\pdfoutput=1

\documentclass[twocolumn, final, amsmath,amssymb,prb,showkeys,superscriptaddress]{revtex4-1}
\usepackage[colorlinks=true, linkcolor=blue,citecolor=blue, urlcolor =blue]{hyperref}

\usepackage{pdfpages} 
\usepackage{graphicx}
\usepackage{bm}
\usepackage[normalem]{ulem}

\usepackage{booktabs}
\usepackage{xparse}
\NewDocumentCommand{\rot}{O{45} O{1em} m}{\makebox[#2][l]{\rotatebox{#1}{#3}}}%

\begin{document}


\title{3D Dirac cone carrier dynamics in Na$_3$Bi and Cd$_3$As$_2$}

\author{G. S. Jenkins}
    \homepage{http://www.irhall.umd.edu}
    \email{GregJenkins@MyFastMail.com}
    \affiliation{Department of Physics, University of Maryland at College park, College Park, Maryland, 20742, USA}
    \affiliation{Center for Nanophysics and Advanced Materials, University of Maryland at College park, College Park, Maryland, 20742, USA}
\author{C. Lane}
    \affiliation{Department of Physics, Northeastern University, Boston, Massachusetts, 02115, USA}
\author{B. Barbiellini}
    \affiliation{Department of Physics, Northeastern University, Boston, Massachusetts, 02115, USA}

\author{A. B. Sushkov}
    \affiliation{Department of Physics, University of Maryland at College park, College Park, Maryland, 20742, USA}
    \affiliation{Center for Nanophysics and Advanced Materials, University of Maryland at College park, College Park, Maryland, 20742, USA}
\author{R. L. Carey}
    \affiliation{Department of Physics, University of Maryland at College park, College Park, Maryland, 20742, USA}
    \affiliation{Center for Nanophysics and Advanced Materials, University of Maryland at College park, College Park, Maryland, 20742, USA}
\author{Fengguang Liu}
    \affiliation{Department of Physics, University of Maryland at College park, College Park, Maryland, 20742, USA}
    \affiliation{Center for Nanophysics and Advanced Materials, University of Maryland at College park, College Park, Maryland, 20742, USA}

\author{J. W. Krizan}
    \affiliation{Department of Chemistry, Princeton University,
Princeton, New Jersey 08544, USA}
\author{S. K. Kushwaha}
    \affiliation{Department of Chemistry, Princeton University,
Princeton, New Jersey 08544, USA}
\author{Q. Gibson}
    \affiliation{Department of Chemistry, Princeton University,
Princeton, New Jersey 08544, USA}

\author{Tay-Rong Chang}
    \affiliation{Department of Physics, National Tsing Hua University, Hsinchu 30013, Taiwan}
\author{Horng-Tay Jeng}
    \affiliation{Department of Physics, National Tsing Hua University, Hsinchu 30013, Taiwan}
    \affiliation{Institute of Physics, Academia Sinica, Taipei 11529, Taiwan}

\author{Hsin Lin}
\affiliation{Centre for Advanced 2D Materials and Graphene Research Centre, National University of Singapore, Singapore 117546}
\affiliation{Department of Physics, National University of Singapore, Singapore 117542}

\author{R. J. Cava}
    \affiliation{Department of Chemistry, Princeton University,
Princeton, New Jersey 08544, USA}

\author{A. Bansil}
    \affiliation{Department of Physics, Northeastern University, Boston, Massachusetts, 02115, USA}

\author{H. D. Drew}
    \affiliation{Department of Physics, University of Maryland at College park, College Park, Maryland, 20742, USA}
    \affiliation{Center for Nanophysics and Advanced Materials, University of Maryland at College park, College Park, Maryland, 20742, USA}

\begin{abstract}
Optical measurements and band structure calculations are reported on 3D Dirac materials. The electronic properties associated with the Dirac cone are identified in the reflectivity spectra of Cd$_3$As$_2$ and Na$_3$Bi single crystals. In Na$_3$Bi, the plasma edge is found to be strongly temperature dependent due to thermally excited free carriers in the Dirac cone. The thermal behavior provides an estimate of the Fermi level $E_F=25$ meV and the z-axis Fermi velocity $v_z = 0.3 \text{ eV} \AA$ associated with the heavy bismuth Dirac band. At high energies above the $\Gamma$-point Lifshitz gap energy, a frequency and temperature independent $\epsilon_2$ indicative of Dirac cone interband transitions translates into an ab-plane Fermi velocity of $3 \text{ eV} \AA$.  The observed number of IR phonons rules out the $\text{P}6_3\text{/mmc}$ space group symmetry but is consistent with the $\text{P}\bar{3}\text{c}1$ candidate symmetry. A plasmaron excitation is discovered near the plasmon energy that persists over a broad range of temperature. The optical signature of the large joint density of states arising from saddle points at $\Gamma$ is strongly suppressed in Na$_3$Bi consistent with band structure calculations that show the dipole transition matrix elements to be weak due to the very small s-orbital character of the Dirac bands. In Cd$_3$As$_2$, a distinctive peak in reflectivity due to the logarithmic divergence in $\epsilon_1$ expected at the onset of Dirac cone interband transitions is identified. The center frequency of the peak shifts with temperature quantitatively consistent with a linear dispersion and a carrier density of $n=1.3\times10^{17}\text{ cm}^{-3}$.  The peak width gives a measure of the Fermi velocity anisotropy of $10\%$, indicating a nearly spherical Fermi surface. The lineshape gives an upper bound estimate of 7 meV for the potential fluctuation energy scale.

\end{abstract}


\maketitle

\section{Introduction}
Topological concepts in condensed matter physics have led to the realization of new states of matter.\cite{bansilrev16, *kane10, *qi11, *Moore11} Ongoing generalizations of topological concepts continue to generate profound discoveries. Many of the new predicted emergent properties have been experimentally confirmed, some analogous to concepts originating in particle physics like the Dirac,\cite{wang12,wang13,xuHasan13,LiuZXShen14,xiongOngCA15,neupane14,liuNature14,jeon14,yi14,borisenkoCava14} Weyl,\cite{wan2011, *burkov11, *weng14,*huang15,*lv15,*xu15,*lvn215} and Majorana fermion.\cite{fu08,*mourik12,* nadj14}
In the condensed matter version, Dirac fermions exist in the valence and conduction bands of 3-dimensional (3D) Dirac semimetals, which touch at a pair of points and disperse linearly away from the nodes. These bands derive from 4-fold degenerate band crossings that are protected against gapping by crystal symmetry. If either crystal inversion or time reversal symmetry is broken, each Dirac node splits into a pair of opposite chirality Weyl nodes, topological objects that act as a source or sink of Berry's phase curvature. This topological band structure effect is analogous to opposite-polarity magnetic monopoles residing at the nodes in momentum space, which fundamentally alter the semiclassical equations of motion and Maxwell's constitutive relations.\cite{zhong15,*goswami13,*zyuzin12,*fujikawa79} Some of the unique properties that may be exploited in potential technological applications include Fermi-arc surface states, chiral pumping effects, and magneto-electric-like effects in plasmonics and optics in the absence of an applied field.\cite{xuHasan13,LiuZXShen14,borisenkoCava14,neupane14,xiongOngCA15,xiongOng15,Hofmann14, Hofmann15,ashby14}

Unlike surface probes like photoemission and tunneling spectroscopy, optical measurements probe bulk band structure and carrier dynamics over a broad range of energy scales. In many ways, optical measurements are ideal probes of the bulk electronic properties of 3D Dirac systems. Sensitive measurement of the free carrier response is possible due to the low carrier densities achievable in Dirac semimetals. The Dirac interband transitions extend down to zero frequency as the carrier density becomes vanishingly small.\cite{hosur12, ashby14} This behavior of the interband transitions gives rise to a logarithmic singularity in the static dielectric constant. The logarithmic divergence, analogous to the ultraviolet divergence encountered in quantum electrodynamics, leads to charge renormalization\cite{Hofmann14, Hofmann15} and screening effects.\cite{skinner14, throckmortonPRB15} Another interesting aspect of 3D Dirac systems is the strong electron-electron interactions characterized by the ratio of the Coulomb to kinetic energy equal to an effective fine structure constant $e^2/(\hbar v_F  \epsilon)$ that is substantially larger than 1 for typical values of the Fermi velocity $v_F$ and dielectric constant $\epsilon$. This behavior is striking since the interaction strength is independent of carrier density, and has been predicted to give rise to plasmaron modes at finite density that could be optically accessible.\cite{Hofmann14, Hofmann15, tediosi2007} Optical probes are also sensitive to predicted signatures of the chiral anomaly as well as the underlying chiral nature of the Weyl states using magneto-optical measurement schemes in zero field.\cite{ashby14,sonspivak13, Hofmann15,goswami15, kargarian15, hofmann16}

Since Na$_3$Bi is highly reactive with air,\cite{kushwaha15} no optical measurements have previously been reported. Providing broadband optical access to samples in a cryogenic environment while protecting them from atmospheric water and oxygen presents substantial obstacles. The high mobility of Cd$_3$As$_2$,\cite{rosenberg59,liangOng15} historically known as a narrow band semiconductor with inverted bands and non-parabolic conduction band, \cite{arush92} attracted many optical studies over the last half century.\cite{turner1961,*haide66,*gelten80,houde86} Only recently have theoretical concepts been developed predicting a pair of Dirac cones\cite{wang13} and subsequent confirmation of their existence by surface probe measurements.\cite{liuNature14,jeon14,yi14,borisenkoCava14,neupane14} Therefore previous optical studies do not report optical effects unique to a Dirac cone except for two very recent optical measurements of Cd$_3$As$_2$. One of these studies reports a nearly constant $\epsilon_2$ in the mid-IR spectral region interpreted as a Dirac cone signature,\cite{neubauer16} while the other is a broadband cyclotron resonance study reporting a linear band structure.\cite{akrap2016}

In this article, the optical spectra of Cd$_3$As$_2$ and Na$_3$Bi are presented together with parallel first-principles band structure calculations. The expected optical signatures and thermal occupation effects in a Dirac cone pair is discussed in section \ref{sec:expectedOptics}. The first optical characterization of Na$_3$Bi is reported and discussed in section \ref{sec:NaBiResults}. In section \ref{sec:CdAsResults}, a peak in reflectivity in  Cd$_3$As$_2$ identifies the onset of Dirac cone interband transitions. A summary of results is presented in section \ref{sec:conclusion}.
\begin{figure}
\includegraphics[width = 7 cm]{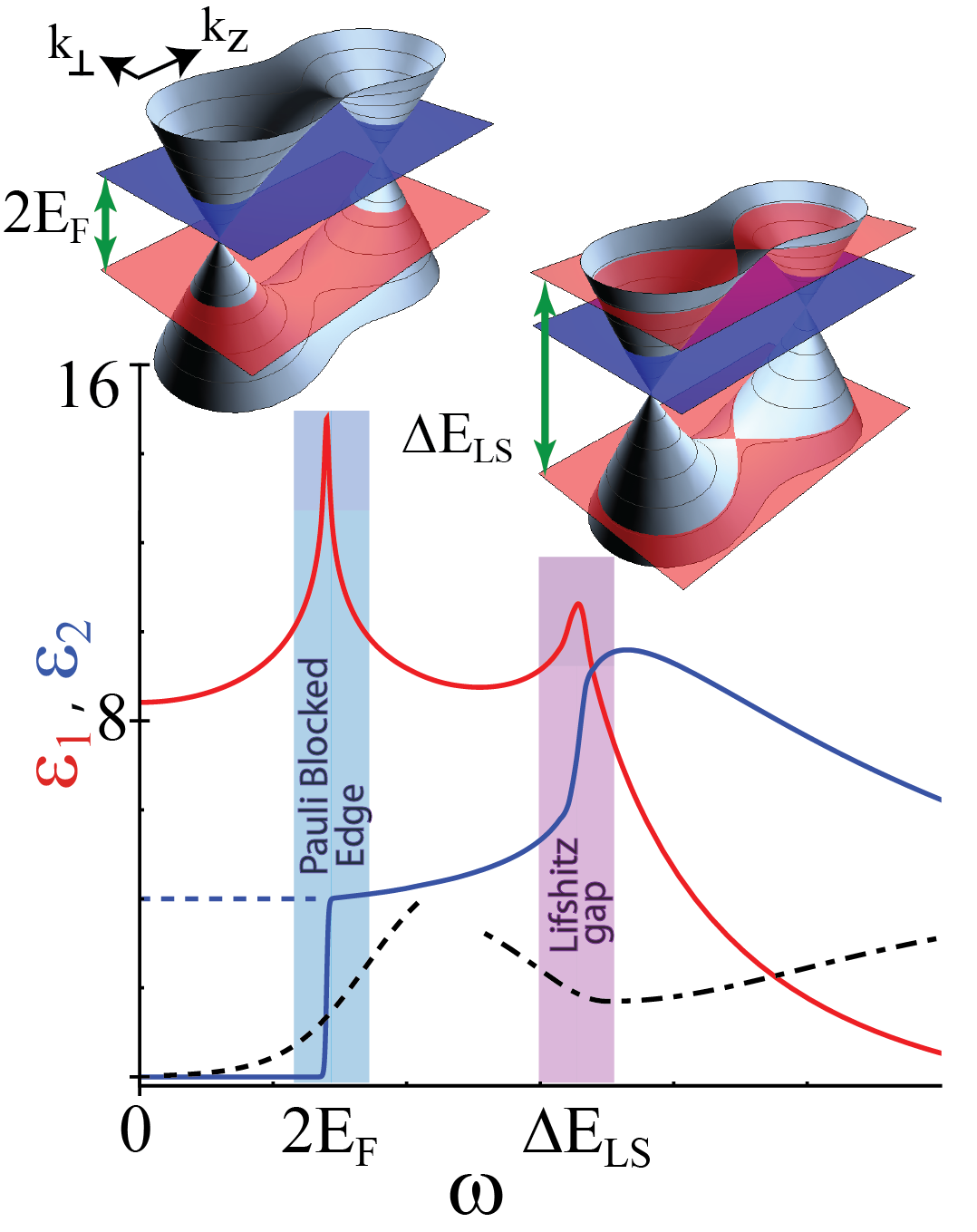}
\caption{\label{fig:ExpectedOpticalSigs}  (a)
The model dispersion of a Dirac cone pair ($k_\perp \equiv k_x=k_y$), which is used to numerically calculate the dielectric function ($\epsilon=\epsilon_1 + i \epsilon_2$), is graphed as solid red and blue lines. The Fermi level (blue plane), $E_F$,  lies in the conduction band between the Dirac point and the saddle-point located at the $\Gamma$-point, midway between the Dirac nodes. Intersections of the red planes with the model dispersions depict the initial and final state energies of optical transitions for two cases: the onset of Dirac cone transitions at the Pauli-blocked edge $\omega=2 E_F$ , and transitions between the two saddle points at the Lifshitz gap energy $\omega=\Delta E_{LS}$. In the graph, the blue-dotted horizontal line is the expected $\epsilon_2$ in the low frequency limit with the Fermi level at the Dirac node predicted by the $k\cdot p$ dispersion\cite{wangPRB12} with anisotropic Fermi velocities, where $v_{z2}=5 \text{ eV} \AA>>v_{z1}$ (see Appendix \ref{app:kdotp}). At finite frequency in the vicinity of the Pauli-blocked edge, a highly anisotropic Fermi surface will broaden the edge as qualitatively depicted by the black-dashed curve of $\epsilon_2$. In the vicinity of the saddle points, band structure calculations show that the dipole transition matrix elements are suppressed, which reduces $\epsilon_2$ as qualitatively shown by the black dash-dotted curve. }
\end{figure}

\begin{figure}
\includegraphics[width=7 cm]{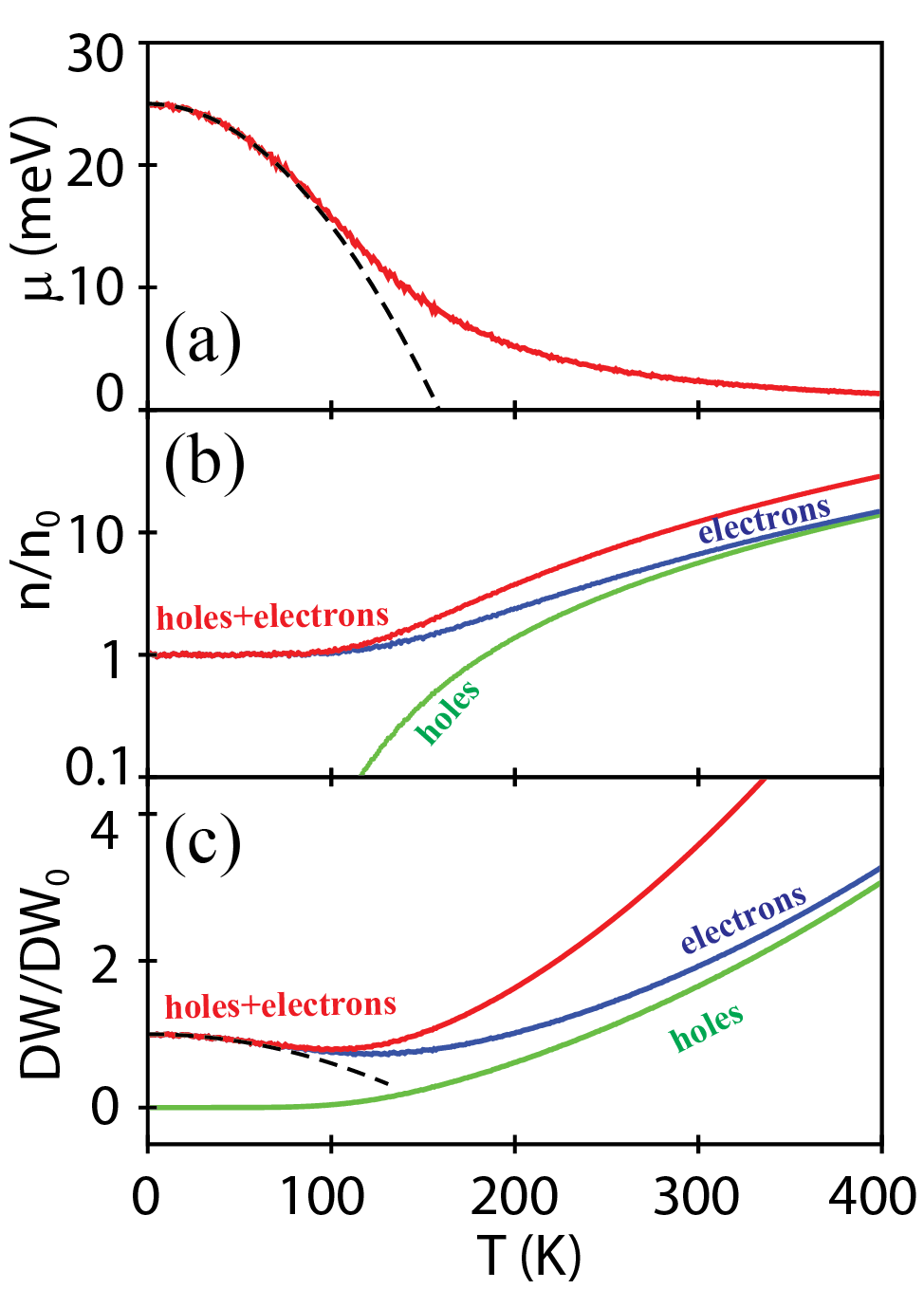}
\caption{\label{fig:ExpectedOpticalThermalSigs}
 A single Dirac cone is presumed to be electron-hole symmetric with a Fermi level $E_F=25 \text{ meV}$ in the conduction band. The chemical potential shown in panel (a) is numerically calculated from the dispersion (red curve) and analytically derived in the low temperature limit (black-dashed curve). The carrier density and Drude weight of thermally excited carriers are shown in panels (b) and (c), respectively, where the contributions from electrons (blue), holes (green), and the sum of the two (red), are plotted. The analytic solution of the Drude weight in the low temperature limit is shown as the black-dashed line in panel (c). }
\end{figure}

\section{expected optical signatures in Dirac cone systems}\label{sec:expectedOptics}

\subsection{Interband transitions}\label{sec:OpticsIdealDiracCones}
In an ideal 3D Dirac cone with the Fermi level at the node, interband transitions occur at all frequencies and give rise to a linear conductivity $\sigma_1 \sim \omega/v_F$ where $v_F$ is the Fermi velocity and $\omega$ is the photon energy.\cite{ashby14} At a nonzero Fermi level, the interband transitions are blocked by carrier occupation below $\omega = 2 E_F$. The lost interband spectral weight below $2 E_F$ gives rise to an equal free carrier (Drude) spectral weight thereby satisfying the f-sum rule.

Since the complex dielectric function is given by $\epsilon = (4 \pi / i \omega ) \sigma$, the Dirac cone interband transition contribution leads to $\epsilon_2=(1/6) N_d \alpha' \Theta(\omega -  2 E_F)$ that is constant above the transition onset, where $N_d$ is the degeneracy of the Dirac cone, $\alpha'=e^2/\hbar v_F$ is the effective fine structure constant, and $v_F$ is the Fermi velocity. The frequency independence of $\epsilon_2$ results from cancellations that occur in Fermi's golden rule between the joint density of states and the dipole transition matrix elements for a linear dispersion.\cite{ashby14,YuCardona}

The Kramers-Kronig transformation of the interband $\epsilon_2$ gives
$\epsilon_1\propto \log{ \frac{\Delta^2 - \omega^2}{(2 E_F)^2 - \omega^2}}$ where $\Delta$ is an energy cut-off defined by the bandwidth. Temperature broadening of the interband transition onset is taken into account by replacing the Heaviside step function in $\epsilon_2$ with a Fermi-distribution function expression, which results in $\epsilon_1\propto \textrm{Re} [\log{ \frac{\Delta^2 - \omega^2}{(2 \mu - \imath \pi T)^2 - \omega^2}}]$ where $\mu$ is the chemical potential and T is the temperature.

For the case of Na$_3$Bi, two Dirac cones are separated by $\delta k_d = \pm 0.1 {\AA}^{-1}$  along the $k_z$ direction. \cite{xuHasan13,LiuZXShen14} The conduction and valence band of the Dirac cones  merge, forming two saddle points at the $\Gamma$-point midway between the nodes as depicted by the idealized dispersion in Fig. \ref{fig:ExpectedOpticalSigs}. The Fermi velocity $v_F \approx 2.5$ eV$\cdot {\AA}$ at each Dirac node is reasonably consistent with photoemission and  transport measurements, and band structure calculations.\cite{xuHasan13,LiuZXShen14,xiongOngCA15,xiongOng15,kushwaha15,wangPRB12,narayan14,cheng14}  For illustration purposes, the Lifshitz gap at the $\Gamma$-point is arbitrarily set at $\sim 4 E_F$.

Considering this dispersion with the Fermi level set at the Dirac node in the limit  $\omega \rightarrow 0$,  the two Dirac cones are well described by the ideal case, so that $\epsilon_2=(1/6) N_d \, \alpha' \approx 4$. This low frequency value is depicted by the dashed-blue line in Fig. \ref{fig:ExpectedOpticalSigs}, providing an estimated scale of the expected interband optical response. This scale also applies to the anisotropic Dirac cone case derived from band structure calculations in the low frequency limit (see Appendix \ref{app:kdotp}). At nonzero values of the Fermi level, the Pauli-blocked edge occurs at $\omega=2 E_F$, giving rise to a step in $\epsilon_2$ and a distinctive cusp-like lineshape in $\epsilon_1$ as shown in Fig. \ref{fig:ExpectedOpticalSigs}.

At higher photon energies, the nonlinearity of the bands along $k_z$ between the nodes become increasingly important. The saddle point region gives rise to a large and rapidly changing joint density of states as well as dipole transition matrix elements that strongly deviate from the linear Dirac case. Both effects should be considered for describing the optical response even though such modeling is numerically difficult.  By ignoring the effects of the transition matrix elements, the effects arising from corrections to the joint density of states can be calculated.\cite{YuCardona}  For this simplified case, a rendering of the features in $\epsilon$ is shown in Fig. \ref{fig:ExpectedOpticalSigs} in the vicinity of the saddle points. We will return below to consider contributions from the dipole transition matrix elements near the saddle points.

 Two main features are thus expected in the optical signal from the Dirac cone interband transitions, one related to the Pauli-blocked edge and the other to the high density of states at the Lifshitz gap energy $\Delta E_{LS}$ at the $\Gamma$-point. The magnitude of the interband contributions to $\epsilon$ is therefore expected to be in the vicinity $\sim 5$ based on reasonable Fermi velocity estimates.

\subsection{Thermal occupation effects}
In Dirac systems with relatively low Fermi level, the temperature dependence of the chemical potential and carrier density can be substantial. These thermal occupation effects can therefore drive observable optical effects.\cite{sushkov2015} The Pauli-blocked edge will thermally broaden, and shift as $\omega= 2 \mu(T)$. The free carrier (Drude weight) response will also change consistent with the f-sum rule.

An analytic form of the chemical potential $\mu$ in the low $T$ limit is  $\mu-E_F=  - \frac{1}{6}  \partial_E \ln[g(E_F)] \,(\pi T)^2 $  where $g$ is the density of states. \cite{ashcroft} Here the band dispersion information is encoded via the derivative of the density of states. An isotropic 3D conduction band where the dispersion is given by $E \propto k^\beta$  (where $\beta=1$ for a Dirac band) results in $ \partial_E \ln[g(E_F)]= E_F^{-1} (3-\beta)/\beta$. The chemical potential is therefore driven away from regions of higher density of states as temperature is increased. The Drude weight $D_W=n e^2/m$ depends on the energy dependence of both the number density $n$  and mass $m$, but for a linear dispersion the energy dependence is given by $D_W=N_d\frac{e^2}{6 \pi^2 \hbar^3}\frac{E_F^2}{v_F}$, where $N_d=4$ is the degeneracy for a Dirac cone pair, and in the low temperature limit we now obtain $\Delta D_W(T)/D_W(0) = -\frac{1}{3}(\frac{\pi T}{E_F})^2$.\cite{throckmortonPRB15}

Numerical solutions for the temperature dependence of the chemical potential, carrier density, and Drude weight are shown in Fig. \ref{fig:ExpectedOpticalThermalSigs}(a-c) for a Dirac cone (see Appendix \ref{app:chempot} and \ref{app:EggFS}). The dispersion is assumed linear with anisotropic velocities, where  $v_z$ can differ from $v_\perp \equiv v_x=v_y$ (resulting in an ellipsoidal or an egg-shaped Fermi surface described in Appendix \ref{app:EggFS}), and the applied electric field is in the x-y plane. The results shown in Fig. \ref{fig:ExpectedOpticalThermalSigs}(a-c) are independent of the velocities and only depend on the Fermi energy that is set to 25 meV. When the chemical potential is within the half-width of the Fermi distribution function ($\pi T/2$) of the Dirac node, copious numbers of additional electrons and holes are thermally excited as shown in Fig. \ref{fig:ExpectedOpticalThermalSigs}(b). The Drude weight involves the sum of responses from holes and electrons, so that in the high temperature limit where the chemical potential is approximately zero and constant, $D_W \propto T^2$. In the low temperature limit, the decreasing $D_W$ with temperature is caused by the decreasing chemical potential $D_W\sim \mu^2 \sim -T^2$.\cite{throckmorton15}  The temperature-dependent Drude weight is therefore nonmonotonic. The minimum demarcates the point where a substantial number of holes and electrons are thermally excited from the valence band, $\mu(T)\sim\pi T/2$.

\section{N\lowercase{a}$_3$B\lowercase{i} Results}\label{sec:NaBiResults}

\begin{figure*}
\includegraphics[width=\textwidth]{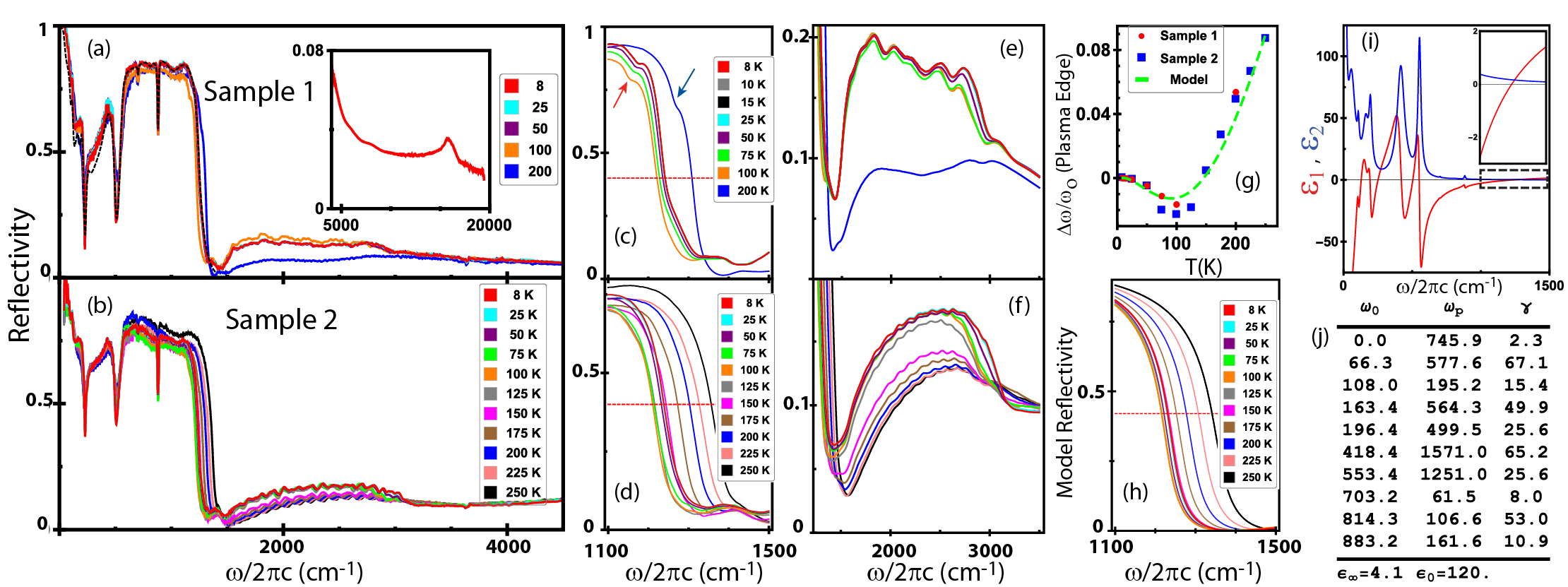}
\caption{\label{fig:Refdataphonons} (a-f) Reflectivity of Na$_3$Bi for two samples are labeled Sample 1 (a,c,e) and Sample 2 (b,d,f). The black-dashed curve in panel (a) is a fit to the 8K spectra below 1400 cm$^{-1}$ using  the Lorentzian oscillator parameters for $\epsilon$ given in panel (j), and shown in graph of panel (i). For Sample 1, a more sensitive detector was used in panels (c) and (e) compared with the broadband measurements reported in panel (a). A strongly temperature-dependent, screened plasma edge is observed in panels (c) and (d). The arrows in panel (c) point to absorptive features (a plasmaron excitation) that tracks the temperature-dependent plasma edge. Panel (h) shows a graph of a reflectivity model based on the parameters in panel (j) except that the free carrier response is replaced by a temperature dependent Drude weight consistent with thermal occupation effects in a Dirac cone. The plasma edge shifts are tracked along the red-dotted lines and summarized in panel (g), where the edge position shifts are normalized to the low temperature value $\omega_0$. The green-dashed model results of the plasma edge shifts are the same as those shown in panel (h) along the red-dotted line except with many more temperatures represented. Panels (e) and (f) show the temperature dependence of the interband transition spectral region.}
\end{figure*}

\subsection{Na$_3$Bi spectra, phonons, and crystal symmetry}

Single crystals of n-doped Na$_3$Bi were prepared as in Ref. \onlinecite{kushwaha15}. All manipulations were performed inside a nitrogen filled glove box to avoid air exposure, including the mounting and sealing of the sample inside a cryostat. The as-grown facets are c-axis (001) oriented.

The normal-incidence reflectivity spectra of two Na$_3$Bi crystals at a set of temperatures are reported in Figs. \ref{fig:Refdataphonons}(a) and (b). The largest two crystals are optically thick (opaque),  accommodate a 2.4 mm and a 1.5 mm diameter aperture, and are labeled as Sample 1 and Sample 2, respectively. The reflection approaches unity at low frequency indicative of a metallic response, and phonon features are observable throughout the far-infrared (FIR) region. The high reflection in the range $650-1200 \text{ cm}^{-1}$ is a reststrahlen band. The screened plasma frequency is near 1300 $\text{cm}^{-1}$ indicated by a sharp edge accompanied by a reflection minimum. Features higher in frequency are due to electronic interband transitions.

The most accurately normalized spectra are from the largest and flattest crystal, so the 8K reflectivity spectrum of sample 1 is fit up to 1400 $\text{cm}^{-1}$ to determine the free carrier and phonon parameters. The model reflectance is generated from the dielectric function $\epsilon = \epsilon_{\infty}+\sum_j {\Omega^2_P}_j / ({\omega_0^2}_j-\omega^2- i \gamma \omega)$  where each Lorentzian oscillator represents a phonon mode with a center frequency $\omega_0$, characteristic width $\gamma$, and strength $\Omega_P$. The free carrier (Drude) response corresponds to $\omega_0=0$ where $2 \pi c \gamma= 1/\tau$ is the inverse lifetime of the carriers, and $\Omega_P^2= 4 \pi D_W$ where $\Omega_P$ is the bare (unscreened) plasma frequency.

The modeled reflectance that best fits the spectrum also incorporates a thin dielectric film on the Na$_3$Bi crystal. The best fit to the reflectivity data of sample 1 was found with a $2 \text{ }\mu\text{m}$ thick dielectric film with an index set to $n=1.9$. The optical path length is consistent with the faint but visibly colored interference patterns observable under magnification from the as-cleaved samples. The thin film model smoothly modifies the photometrics over a very broad range, with a weak periodic Fabry-Perot-like etalon period of $1200 \text{ cm}^{-1}$. When the thin film is removed from the model, the resulting spectrum better resembles the spectrum of sample 2 in Fig. \ref{fig:Refdataphonons}(b). The thin dielectric is therefore attributed to a surface layer on sample 1 which is inconsequential to the results presented. The fit to the reflectivity spectrum is shown by the dashed-black curve in Fig. \ref{fig:Refdataphonons}(a), and the bulk Na$_3$Bi parameters and associated dielectric function are reported in Figs. \ref{fig:Refdataphonons}(j) and (i), respectively.

The observed phonon spectrum is important since the number of IR active phonon modes relates to the crystal symmetry. The ground state of Na$_3$Bi is currently contentious.\cite{cheng14} The strongest observed phonons at 418 and 553 $\text{cm}^{-1}$, which give rise to the broad reststrahlen band, are a factor of two larger than the predicted highest phonon frequency from our \textit{ab initio} band structure calculations that agree with earlier studies.\cite{cheng14} Three candidate crystal symmetries are analyzed using point group analysis and the number of allowed acoustic, IR active, and Raman active phonons are reported in Appendix \ref{app:PGAnalysis}.

A recent x-ray study reports that Na$_3$Bi is in the hexagonal space group P6$_ 3$/mmc.\cite{kushwaha15} The unit cell consists of two formula units with a Na(1)-Bi honeycomb structure separated by interstitial Na(2) atoms. The number of expected phonons is therefore 24, of which 2 are expected to be IR active in the ab-plane. This is inconsistent with the 9 minimum observable oscillators reported in Fig. \ref{fig:Refdataphonons}(j) necessary to describe our data, which rules out the P6$_ 3$/mmc symmetry.

A recent \textit{ab initio} calculation shows that the P$\overline{\text{3}}$c1 and P6$_3$cm ground states are $\sim4$ meV lower than the P6$_3$/mmc structure. All three point group symmetries produce nearly the same x-ray diffraction pattern and similar Dirac cone bands.\cite{cheng14, wang12} The P$\overline{\text{3}}$c1 and P6$_3$cm structures, however, have a distorted Na-Bi honeycomb resulting in additional inequivalent Na Wykoff sites. The unit cell therefore increases from two formula units to six, and the number of phonon modes triples. Eleven infrared active phonons in the ab-plane are expected from point group analysis in both buckled-hexagonal-plane symmetries. The optical spectrum is fit well with the minimum of 9 phonon oscillators, but some are unusually broad which could imply multiple closely-spaced phonons. The optical data appear consistent with either the P$\overline{\text{3}}$c1 or  P6$_3$cm structure.

However, the P6$_3$cm symmetry has no center of inversion and therefore cannot be a Dirac semimetal, but would rather split into a Weyl state system with four nodes. There is no evidence from surface probe measurements that this is the case. Furthermore, numerical calculations show that the P6$_3$cm symmetry (as well as the P6$_3$/mmc structure) may be unstable due to the existence of imaginary phonons.\cite{cheng14} Therefore, Na$_3$Bi likely belongs to the P$\overline{\text{3}}$c1 spacegroup.

The Drude fit parameters are determined by the low frequency response and the plasma edge feature. Some uncertainty is introduced since the zero-frequency Lorentzian is not sufficiently distinguishable from low frequency bismuth phonons. Reasonable fits to the data give a range of Drude parameters, where $\gamma < 15 \text{ cm}^{-1}$ and $500 \text{ cm}^{-1} <\Omega_P<1000 \text{ cm}^{-1}$. The Fermi level is estimated from the plasma frequency using a model dispersion. A Dirac cone model is described in Appendix \ref{app:EggFS} that produces an elongated egg-shaped Fermi surface. This shape approximates the Fermi surface produced by a more realistic dispersion derived from a $k\cdot p$ model with parameters that fit the Dirac cone bands obtained from first-principles numerical band structure calculations.\cite{wangPRB12} The Fermi level is then estimated by $E=\sqrt{3\pi\hbar^3 v_{z1}/N_d} \Omega_P$ where the degeneracy $N_d=4$ and $v_{z1}$ is the slower of the two velocity roots along the c-axis. For $v_{z1}= 0.5 \text{ eV} \AA$ as measured by photoemission (ARPES),\cite{xuHasan13}  the Fermi energy ranges from $16 \text{ meV} <E_F< 34 \text{ meV}$, and is $25\text{ meV}$ for the  Drude best fit parameter $\Omega_P=746 \text{ cm}^{-1}$.

The static dielectric constant is $\epsilon_0=120^{+10}_{-30}$. The uncertainty is based upon the uncertainty in $\Omega_P$ and therefore the uncertainty in the strength of the low frequency phonons.

\subsection{Pauli-blocking and Lifshitz gap}
Since Na and Bi are relatively heavy atoms, phonon features are relegated to low frequency, well below the measured plasma edge, as verified by our band structure calculations.\cite{cheng14} Considering the estimate of the Fermi level and consulting the band structure calculations in Fig. \ref{fig:bandstructure}(a-d) (our results for the three candidate symmetries verify those of references \onlinecite{wangPRB12} and \onlinecite{cheng14}), a conservative estimate of the spectral region where a Pauli-blocked edge may be found is between $300$ and $1500 \text{ cm}^{-1}$. Nearly this entire region is within the reststrahlen band where the reflectivity is extremely sensitive to small features in $\epsilon$ on the scale expected by a sharp Pauli-blocked edge $\sim 5$,  as demonstrated by the phonon features in the reflectivity located at $700$ and $880 \text{ cm}^{-1}$ produced by much smaller associated $\epsilon$ features shown in Fig. \ref{fig:Refdataphonons}(i). Furthermore, the steep slope of the plasma edge and the deep minimum in the reflectivity just above the plasma edge in the vicinity of $1300 \text{ cm}^{-1}$, where $\epsilon_1\approx0$ and therefore $R_{min}\approx (\epsilon_2/4)^2$, requires $\epsilon_2<1$. An onset of Dirac cone interband transitions anywhere below $1300 \text{ cm}^{-1}$ is expected to contribute a much larger $\epsilon_2$.

No discernable features in the reflectivity spectra resemble the expected features from a Pauli-blocked edge or Lifshitz gap shown in Fig. \ref{fig:ExpectedOpticalSigs}. Band structure calculations show that the assumptions that led to these expectations must be modified. The large anisotropy of the Dirac cone, as demonstrated along the $k_z$ direction ($\Gamma$ -$A$)   in Fig. \ref{fig:bandstructure}(c), gives rise to a wide range of interband transition onset frequencies for a nonzero Fermi level. The Pauli blocked edge therefore becomes broadened, as diagrammatically represented by the black-dashed line in Fig. \ref{fig:ExpectedOpticalSigs}. Furthermore, band structure calculations show that dipole transition matrix elements are strongly modified in the vicinity of the saddle points at $\Gamma$.  The Dirac cone bands in Fig. \ref{fig:bandstructure}(c) have $s$ and $p$ orbital character with a strength proportional to the size of the red dots. Allowable dipole Dirac interband transitions therefore must involve $s\leftrightarrow p$ transitions. The Dirac cone bands along $\Gamma-A$ have $p$ orbital character, but only one of the Dirac bands has  $s$ orbital character and it is strongly suppressed as the $\Gamma$-point is approached. The large joint density of states at the $\Gamma$-point that gave rise to the sharp increase in $\epsilon_2$ in Fig. \ref{fig:ExpectedOpticalSigs} is strongly modified by the diminution of the matrix elements (see the black dotted-dashed line in Fig. \ref{fig:ExpectedOpticalSigs}).

\subsection{Thermal occupation effects and electronic transitions in the Dirac cone}
\subsubsection{Plasma edge and Drude weight temperature dependence}

\begin{figure*}
\includegraphics[width=\textwidth]{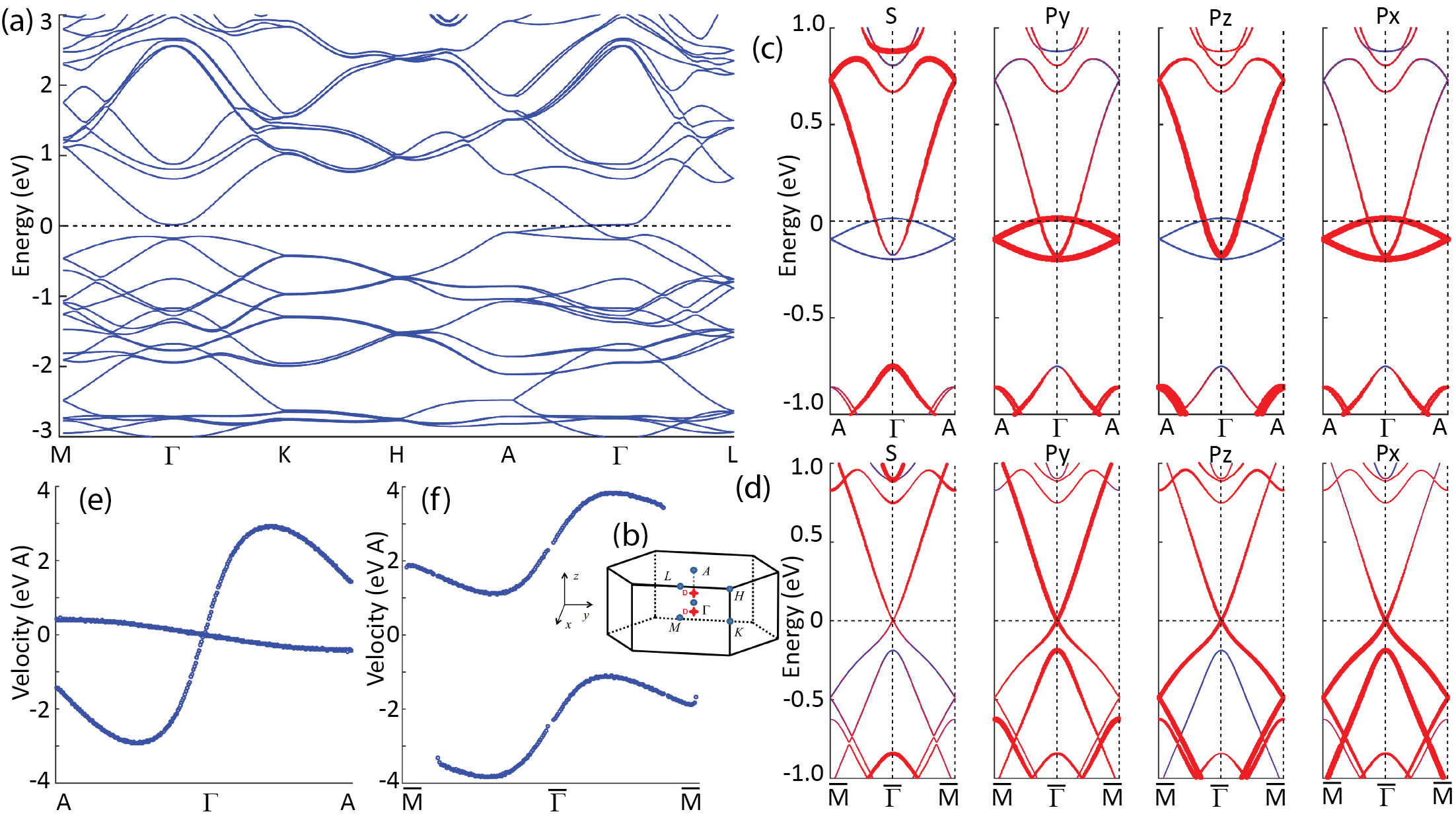}
\caption{\label{fig:bandstructure} (a) The calculated band structure of Na$_3$Bi is shown for the $P\bar{3}c1$ space group. Results are very similar for the $P6_3cm$ structure. (b) The Brillouin zone for the crystal structure in (a) is depicted with Dirac nodes marked by the two red points along $\Gamma-A$. (c,d) The projected orbital-characters of the bands are shown for $s$, $p_y$, $p_z$, and $p_x$ orbitals along the $\Gamma-A$ momentum direction as well as through the Dirac node parallel to the $\Gamma-M$ direction, denoted by $\bar{\Gamma}-\bar{M}$. Bands are plotted as blue lines, overlayed by dotted red lines with thickness proportional to the weight of the orbital character. The orbital character of the bands along the $\bar{\Gamma}-\bar{K}$ direction is very similar to that along $\bar{\Gamma}-\bar{M}$. Panel (c) shows the lack of s-orbital character of the Dirac cone heavy Bi-like band as well as the lighter Na-like band in the vicinity $\Gamma$, which causes the optical transition matrix elements associated with the Lifshitz gap region to be suppressed. (e-f) Fermi velocities of the two Dirac bands are plotted along $\Gamma-A$ and $\bar{\Gamma}-\bar{M}$ for the $P\bar{3}c1$ space group; velocities for the $P6_3cm$ structure are identical. Velocity plots along $\bar{\Gamma}-\bar{K}$ and $\bar{\Gamma}-\bar{M}$ are similar.}
\end{figure*}

Although the Pauli-blocked edge and the Lifshitz gap optical features are complicated by band structure anisotropy and transition matrix elements, the nonmonotonic temperature dependence of the plasma edge summarized in Fig. \ref{fig:Refdataphonons}(g) encodes Dirac cone information. The strength of the zero frequency oscillator in the dielectric function relates to the Drude weight, $D_W=\Omega_P^2/4\pi$. A decrease in Drude weight shifts the zero of $\epsilon_1$, and therefore the plasma edge, to lower frequency. The resemblance between the temperature dependence of the plasma edge in Fig. \ref{fig:Refdataphonons}(g) and of the Drude weight in Fig. \ref{fig:ExpectedOpticalThermalSigs}(c) suggests that the plasma edge shifts are caused by thermal occupation effects in the Dirac cone. As mentioned previously, the results of Fig. \ref{fig:ExpectedOpticalThermalSigs}(a-c) are independent of Fermi velocity for a linear dispersion, even for a Dirac cone with anisotropic velocities, and depend only on the Fermi level. The minimum frequency of the plasma edge in Fig. \ref{fig:Refdataphonons}(g) occurs at $T\approx 100$K. Assuming these shifts are caused by the temperature dependent Drude weight, the Fermi level is estimated to be $E_F=25 \text{ meV}$ since this value gives rise to a minimum in $D_W(T)$ at 100K.

This connection between thermal occupation effects in the Dirac cone that drive the Drude weight temperature dependence and the plasma edge shifts is verified by the quantitative agreement of the reflectivity model results shown in Fig. \ref{fig:Refdataphonons}(h). The temperature dependent Drude weight of Fig. \ref{fig:ExpectedOpticalThermalSigs}(c) with $\Omega_{P0}=950 \text{ cm}^{-1}$ is substituted into the complex dielectric function that includes the phonons reported in Fig. \ref{fig:Refdataphonons}(j) (with the parameter $\epsilon_\infty$  increased by 10 percent) and the reflectivity calculated. Utilizing the results of Appendix \ref{app:EggFS} that show $E_F = \sqrt{3 \pi \hbar^3 v_{z1}/N_d} \Omega_P$ and substituting this value of $\Omega_{P0}$ and $E_F=25$ meV, the slow root of the dispersion which physically corresponds to the conduction band between the nodes is found to be $v_{z1}\approx 0.3 \text{ eV}\AA $. This is a very reasonable number since $v_{z1}\sim v_\perp/10$ as shown by band structure results in Fig. \ref{fig:bandstructure}(e) and ARPES measurements.\cite{xuHasan13,LiuZXShen14} Despite some subtle differences between the measured temperature dependence of the plasma edge of the two samples in Fig. \ref{fig:Refdataphonons}(c) and (d), the model results in Fig. \ref{fig:Refdataphonons}(h) agree extremely well.

The temperature dependence of $\mu$ or $D_W$ ideally contains information associated with the large density of state region at the saddle point as well as the degree of electron-hole asymmetry of the Dirac bands, both of which the model neglects. For example, if the Fermi energy were in the vicinity of the conduction band saddle point where the density of states rapidly increases, the factor $ \partial_E \ln[g(E_F)]$ (in the expression for $\mu(T)$) would be larger than the linearly dispersing value of $2/E_F$. The increase of this factor  would cause the chemical potential to decrease more quickly with temperature than a linear dispersion. As a result, a discrepancy bewteen the model rate of decrease of the plasma edge and data would be expected. Along this line, the discrepancy between the model and Sample 2 at low temperatures could be taken as evidence that the Lifshitz point is in the vicinity of 25 meV above the Dirac point. In principle, a low temperature characterization of $\mu(T)$ or $D_W(T)$ could be used to discern the temperature dependence of the $T^2$ coefficient and therefore determine the Lifshitz transition energy in the density of states in relation to the Fermi level, but the exercise requires many more than four or five low temperature data points (below 100K).

As mentioned already,  the calculations leading to Fig. \ref{fig:ExpectedOpticalThermalSigs}(a-c) assume electron-hole symmetry. In a more realistic Dirac cone pair system with asymmetric saddle points such that $|E_{LS}^{CB}|<<|E_{LS}^{VB}|$, the assumption applies near the Dirac point where linear approximations are valid. In this case, a valence or conduction band Fermi pocket within $\pm|E_F|$ has the same size and shape.  However, the assumption breaks down when the chemical potential and thermal half-width approach the Lifshitz energy $\mu(T) + \pi T/2\sim E_{LS}^{CB}$. The low temperature consequences were discussed in the previous paragraph. At high temperature, the chemical potential will be pushed below the Dirac node. A numerical calculation with electron-hole asymmetry such that  $E_{LS}^{CB}\sim 30 \text{ meV}=(1/2) |E_{LS}^{VB}|$ and $E_F=25 \text{ meV}$ results in a chemical potential which crosses zero at about 150K reaching $-10$ meV at 300 K. This effect on $\mu(T)$ lowers the temperature of the Drude weight minimum a small amount, where $\mu(T)=\pi T/2$ gives $T=90K$, but does not significantly effect the high temperature Drude weight since the thermal width becomes substantially larger than the chemical potential. The upshot is that even fairly large asymmetries between valence and conduction bands do not appreciably modify the quantitative conclusions of the thermal analysis presented in this section.

\subsubsection{Interband transitions and thermal occupation of the Dirac cone saddle point}

A strong temperature dependence is observed over the interband transition region between $1500 $ and $3000 \text{ cm}^{-1}$. The reflectance over this entire spectral region continually decreases with temperature, but precipitously drops in the temperature range between $125$K and $150$K.

\begin{figure}
\includegraphics[width=8 cm]{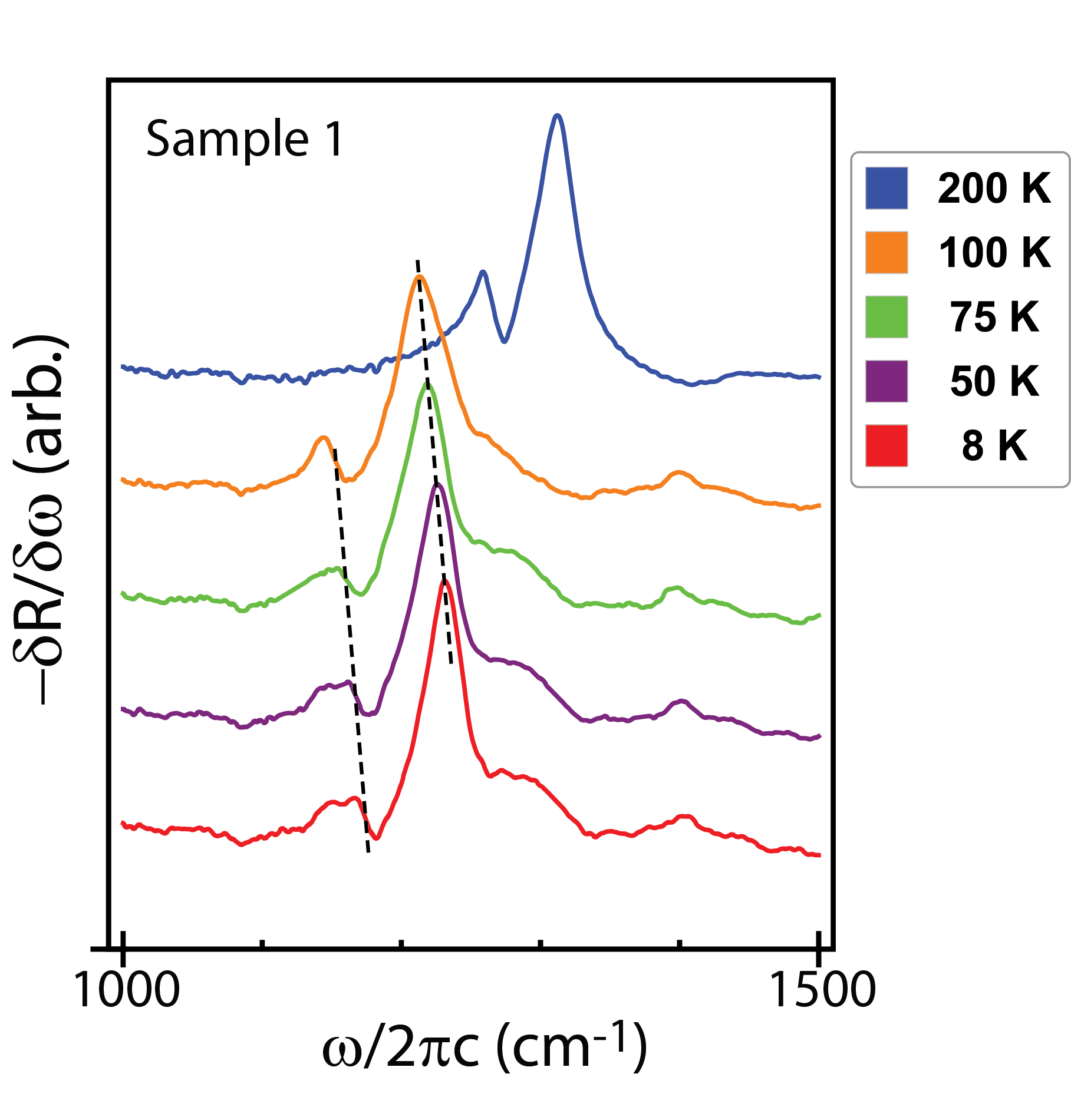}
\caption{\label{fig:plasmaron}
The reflectance $R$ shown in Fig. \ref{fig:Refdataphonons}(c) for Sample 1 in the vicinity of the plasma edge at a set of temperatures is plotted as $\partial R/\partial\omega$. The curves are vertically offset for clarity. The highest peak of each plot is associated with the plasma edge. A dip feature in Figs. \ref{fig:Refdataphonons}(c) and  \ref{fig:Refdataphonons}(d) below the plasma edge frequency is present in both samples, although it is much sharper in sample 1 as highlighted by the arrows in Fig. \ref{fig:Refdataphonons}(c). The feature manifests as a peak-dip structure in $\frac{\partial R}{\partial \omega}$, like a side lobe to the plasma edge peak, that tracks the plasma edge as it moves with temperature. The two black-dashed parallel lines are guides for the eye that show that the peak-dip plasmaron feature tracks the plasma edge up to $100$ K, and clearly persists at $200$ K.}
\end{figure}

As mentioned previously, three crystal structures  considered in this study have nearly the same ground state energy to within a few meV.\cite{cheng14} This suggests that a phase change may occur as a function of temperature. However, the IR active phonons shows no anomalous behavior. Also, band structure calculations were performed for the three candidate crystal symmetries in which the lattice spacing was varied to simulate temperature changes. No discernable changes in the electronic structure or orbital characters were identified that correlated to the observed behavior.

Thermal occupation effects of a band with a large density of states within $\pi 150K/2\sim 20$  meV of the chemical potential provides a plausible explanation of the observed behavior. At these high temperatures, the chemical potential is expected to be near the Dirac point. Based on the band structure calculations in Figs. \ref{fig:bandstructure}(a-d), the only conduction band that is in the vicinity of $20$ meV of the Dirac node is the Dirac cone conduction band saddle point, which has only $p$-orbital character.

A candidate valence band with $s$-orbital character exists at the $\Gamma$-point, but lies $\sim750$ meV below the Dirac node as shown in Fig. \ref{fig:bandstructure}(c). Band structure calculations show that the energy of this band is very sensitive to the spin-orbit coupling strength. Decreasing the spin-orbit coupling by a factor of two does not significantly alter the Dirac cone bands, but pushes the s-band up in energy by about a factor of two. The optical results together with band structure calculations may therefore provide a sensitive method to determine the spin-orbit coupling strength.

In this picture, transitions at low temperature between this $s$-character valence band and the $p$-character Dirac cone conduction band give rise to allowable transitions in the vicinity of the $\Gamma$-point with a large joint density of states, provided that $E_F<E_{LS}^{CB}$. As the temperature is raised and the chemical potential lowers toward the Dirac point, these transitions remain active until the thermal broadening is large enough that a copious number of carriers occupy the conduction band saddle-point region. The thermal occupation of the final states at high temperatures will therefore suppress these interband transitions.

The temperature dependence of these interband transitions is only appreciable up to $\sim 3000 \text{ cm}^{-1}$ since, away from the $\Gamma$-point in the Dirac conduction band along the $k_\perp$ direction, the final state energy of interband transitions rapidly increases above the scale associated with thermal occupation effects.

\subsection{Dirac cone transitions above the Lifshitz energy}
The higher energy transitions above $3000 \text{ cm}^{-1}$ are larger than the Lifshitz gap energy where the Dirac cone pair merges into a single Dirac cone. Over the spectral range $\sim3000-6000 \text{ cm}^{-1}$, $\epsilon_2=1.5\pm0.2$  is frequency and temperature independent, which is derived from fitting the reflectivity using a Kramers-Kronig constrained variational dielectric function.\cite{kuzmenkoVDF2005} Since $\epsilon_2=(1/6) N_d \alpha'$ where $N_d=2$ for a single Dirac cone,  a reasonable Fermi velocity of $v_F \approx 3 \text{ eV }\AA$ in the ab-plane is attained consistent with other measurements of $v_\perp$.\cite{xuHasan13,LiuZXShen14,kushwaha15,xiongOng15}

\subsubsection{Plasmaron feature}
Fig. \ref{fig:Refdataphonons}(c) shows a dip feature, indicated by the arrows, about $60 \text{ cm}^{-1}$ below the plasma edge, which tracks the temperature dependence of the plasma edge.  This tracking behavior is more clearly observed by taking the derivative $\partial R/\partial \omega$ shown in Fig. \ref{fig:plasmaron}. The low temperature lineshape of the dip feature in reflectivity  is reproduced by adding a very small Lorentzian absorption to the total dielectric function, which has a characteristic width $\gamma = 40 \text{ cm}^{-1}$ and strength $\Omega_P=50  \text{ cm}^{-1}$ resulting in a small peak value of only $\sim 0.05$ in $\epsilon_2$. Such a tiny absorptive feature is observable only because the total $\epsilon$ is small near the plasma edge.

Sample 2 shows similar behavior in Fig. \ref{fig:Refdataphonons}(d), but the suppression of the reflectivity just below the plasma edge is much broader (as with nearly all the features of Sample 2 in comparison with Sample 1), and appears as a broad sideband-shoulder in $\partial R/\partial \omega$ instead of a clear peak-dip feature.

The observation of an absorption feature that tracks the ab-plane plasma frequency strongly suggests a plasmon-coupled excitation that is electronic in origin. A possible excitation is a charge that couples to the plasmon density modes,\cite{lundqvist67} called a plasmaron excitation, which has recently been predicted in 3D Dirac systems: at a finite value of the Fermi level, the Coulomb interaction induces satellite quasiparticle peaks in the spectral function, which form sidelobes off the main quasiparticle branch.\cite{Hofmann14, Hofmann15}

Plasmaron modes must be excited by a longitudinal field component. A  scattering processes is required that induces the longitudinal mode that can then couple to the c-axis plasmon. Such a process has been observed in similar optical measurements on  bulk bismuth crystals,\cite{tediosi2007,armitageBi10} although the mechanism is far from clear: impurity scattering \cite{gerlachPlasmaron74} and an electron-hole decay scenario has been proposed without reaching a definitive conclusion.\cite{tediosi2007,armitageBi10}

Optically excited plasmaron excitations in 3D materials have rarely been observed, which makes the observation in a 3D Dirac cone system particularly interesting. In the case of elemental bismuth, a plasmaron excitation is observed at a higher energy than the plasmon mode.\cite{tediosi2007}.  For Na$_3$Bi, the c-axis plasmaron excitation is observed below the ab-plane plasmon energy. Therefore, the c-axis plasmon must be lower in energy than the ab-plane plasmon.

The plasmon energy is determined by the pole in $1/\epsilon_z$ and therefore it involves a sum of many contributing terms: free carrier (Drude) response, strength and number of IR active phonon modes, and the high energy interband transitions that cumulatively determine the value of $\epsilon_\infty$. The strength of the c-axis phonons and $\epsilon_\infty$ is not currently known, but can be easily determined optically with an appropriately oriented crystal. What is known is that the Drude weight is smaller for an electric field along the c-axis since the Fermi velocity is smaller than $v_\perp$, and the number of IR active phonons along the c-axis is substantially less than in the ab-plane (see Appendix \ref{app:PGAnalysis}). Both effects would tend to decrease the c-axis plasmon frequency below the ab-plane plasma edge.

Clear evidence of a collective plasmon-electronic excitation in bismuth and now in the 3D Dirac system Na$_3$Bi has been found. Na$_3$Bi and elemental bismuth share many characteristics, such as a Dirac-like (L point) conduction band that has a high Fermi velocity and a small associated Fermi surface, carrier density, and Fermi wavevector.  These observations suggest that collective plasmon-coupled excitations are perhaps more ubiquitous, and open up the possibility of further investigating such collective modes in the various types of Weyl and Dirac systems.

\begin{figure}
\includegraphics[width=\columnwidth]{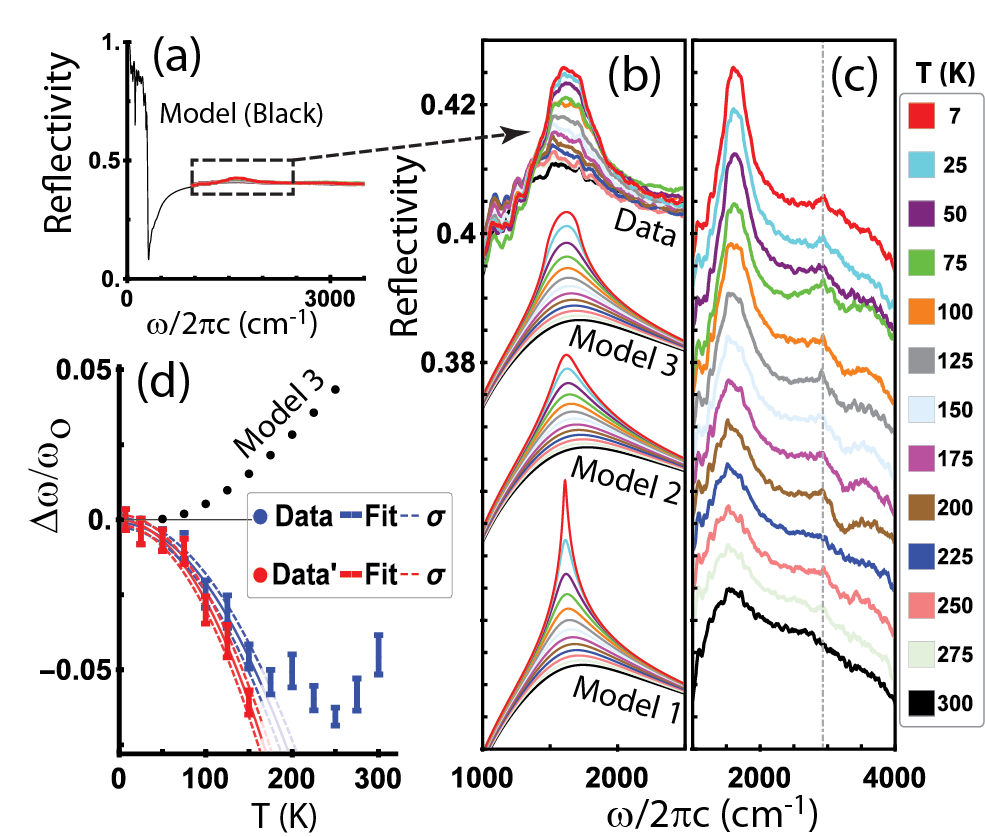}
\caption{\label{fig:CdAs}  (a) The measured mid-IR data (solid colors) is shown with the modeled reflectivity (black) that includes the fitted phonon parameters from Ref. \onlinecite{houde86}. (b) An expanded view of the peak in reflectivity due to the onset of Dirac cone interband transitions and the modeled reflectivities (offset for clarity). Model 1 includes only thermal effects. Contributions to the width in addition to the thermal effects include potential fluctuations, shown in Model 2, or continuum of interband transition onset energies, shown in Model 3. (c) The same data is shown over a broader spectral range, offset for clarity, with a temperature independent feature demarcated by the gray dotted line. (d) The temperature dependent peak positions are fit and plotted relative to the 7 K value $\omega_0$, expressed as $\Delta \omega/\omega_0$, for the data (blue dots), Model 3 in panel (b) (black dots), and the corrected data taking into account the changing slope of the background (red dots). Error bars represent $\pm \sigma$, a standard deviation, generated from fits to the derivative of the peak. Also shown are quadratic fits (solid lines) and $\pm \sigma$ confidence intervals (dotted lines) to the data and corrected data for $T \leq 150$ K.}
\end{figure}

\section{ C\lowercase{d}$_3$A\lowercase{s}$_2$ spectra and Pauli-blocked edge}\label{sec:CdAsResults}

Cd$_3$As$_2$ n-type single crystals were prepared as in Ref. \onlinecite{aliCava14}, and the facet was oriented normal to [112]. The largest crystal accommodates a 0.4 mm aperture and is opaque. A continuous scan FTIR spectrometer measured normally incident reflection.

The small size of the crystal limited throughput power, which precluded measurements in the FIR spectral region. The mid-IR data is reported in Fig. \ref{fig:CdAs}(a-c) at a set of temperatures. A strong temperature-dependent peak in the vicinity of 1650 $\text{cm}^{-1}$ is identified as the Dirac cone Pauli-blocked edge. Band structure calculations and surface probe measurements indicate that other bands do not contribute at such relatively low energies.\cite{wang13, jeon14,liuNature14,neupane14,borisenkoCava14}The peak in the low temperature data implies a Fermi level in the vicinity of $\omega/2 \sim 100$ meV.

Surface tunneling microscopy (STM) measurements and \emph{ab initio} calculations indicate that the Lifshitz gap energy is only $\sim 40$ meV.\cite{jeon14,wang13}  In the scenario where the Fermi energy is much larger than the Lifshitz gap, the Dirac cone pair merges into a single Fermi pocket. ARPES, STM, and transport measurements indicate that the bands appear very linear in this regime\cite{jeon14,liangOng15,liuNature14,neupane14,borisenkoCava14} with nearly isotropic velocity\cite{jeon14,liangOng15} with a single large-Dirac-cone-like dispersion .

We consider a model of reflectivity derived from a dielectric function that includes  contributions from phonons, ideal Dirac cone interband transitions (where $N_d=2$, $E_F=100$ meV, and\cite{liangOng15} $v_F = c/322$), and a Drude weight consistent with the interband transition parameters given by $\Omega_P^2 / (4 \pi)= \frac{2}{3 \pi^2 \hbar^3} \frac{E_F^2}{v_F}$. FIR reflectivity data from Ref. \onlinecite{houde86} is fit to derive the phonon parameters. The model is shown in Fig \ref{fig:CdAs}(a) with the mid-IR reflectance data superimposed. The plasma edge is below our measured frequency range due to limitations in throughput power as a result of the small size of the sample.

By modeling several lineshape broadening effects and comparing to the data, the origin of the distinctive lineshape can be determined. The results of three different models are compared with the data in Fig. \ref{fig:CdAs}(b). Model 1: Thermal effects broaden the Pauli-blocked edge step in $\epsilon_2$ via a Fermi distribution function, which modifies $\epsilon_1$ via the Kramers-Kronig relations. The resulting cusp-like peak in the reflectivity that is dominated by the logarithmic divergence in $\epsilon_1$ is much too narrow to account for the data. Model 2: Gaussian potential fluctuations are added into the step of $\epsilon_2$ in conjunction with thermal broadening. The best match to the width of the 7K reflectivity peak is given by an amplitude potential fluctuation (RMS) $\Gamma_{\text{rms}} \sim 26$ meV. The resulting characteristic lineshape is very different from the data. Notably, an estimate of the potential fluctuations in Cd$_3$As$_2$ given in Ref. \onlinecite{skinner14} is substantially smaller, $\Gamma_{\text{rms}} \sim 4$ meV (using $E_F=100$ meV, $\epsilon_0 = 70$, $v_F$=c/322, $N_d$=2, and an assumed charged impurity density equal to the carrier density). Model 3: Anisotropies between the conduction and valence bands can result in a continuum of interband transition onset frequencies. To model this, the expected step height in $\epsilon_2$ for an ideal Dirac cone is divided into a series of equal step heights separated by equal frequency spacings. Each step is thermally broadened. Model 3 used in Fig. \ref{fig:CdAs}(b) was generated with 20 steps over a frequency range of $125 \text{ cm}^{-1}$.

The strong resemblance of Model 3 to the distinctive lineshape and thermal dependence of the data indicates a continuum of onsets in the Dirac cone over an energy range $\Delta \omega_{\text{onset}}\approx15$ meV.  Incrementally adding in potential fluctuation broadening effects into Model 3 gradually evolves the lineshape towards Model 2, but also tempers the rate of decrease of the low temperature peak heights to better resemble the data. The low temperature lineshapes markedly begin deviating from the data at $\Gamma_{\text{rms}}= 7$ meV, and become untenable by $\Gamma_{\text{rms}}= 10$ meV, which sets a hard upper bound. These values are in reasonable agreement with the theoretical estimate, $\Gamma_\text{rms}\sim 4$ meV.\cite{skinner14}

The spread in Dirac cone interband transitions $\Delta \omega_{\text{onset}}$ is caused by velocity anisotropy of the Dirac bands.  Using the ellipsoidal Fermi surface described in Appendix \ref{app:EggFS}, this energy spread translates into a $10\%$ variation of velocity, a very small degree of anisotropy, and therefore a nearly spherical Fermi surface. This agrees with STM, SdH, and recent cyclotron resonance results that show a nearly isotropic Fermi surface at high Fermi levels well above the Lifshitz gap.\cite{jeon14,liangOng15,akrap2016}

A shift of the peak towards lower frequency with temperature is driven by the chemical potential. This relationship is derived in Appendix \ref{app:chempot} in the low temperature limit.  The relative shift of the peak position normalized to the low temperature value depends only on the Fermi level, where $\delta \omega/\omega_0=-\frac{1}{3}(\frac{\pi T}{E_F})^2$ for a linear dispersion, which is measured by ARPES and tunneling microscopy.\cite{jeon14,liuNature14,neupane14,borisenkoCava14}. The peaks are fit to determine the center frequency. The results are plotted in Fig. \ref{fig:CdAs} (d) as blue dots with error bars. The temperature dependence is fit for $T\leq150$ K to the expected quadratic form (blue solid plot) with confidence intervals as dashed lines. Using the $T^2$ fit coefficient yields $E_F = 111\pm 4$ meV.

Since the experimental peaks reside on a smooth non-constant background, the peak positions are slightly skewed as a function of temperature. To estimate these corrections, the peaks of Model 3, where the center of the Pauli-blocked edge was set to a constant $\omega_0$ for all temperatures, are fit using the same procedure as the experimental data. The centers of peak positions determined in this way are plotted in Figure \ref{fig:CdAs} (d) (black dots), appearing temperature dependent as the peak thermally broadens. These relatively small corrections to the peak positions are subtracted from the experimentally determined positions and reported as red dots and error bars. This corrected dataset is fit as before yielding $E_F = 96\pm 3$ meV, and a carrier density of $n=1.3 \times 10^{17}\text{ cm}^{-3}$. This is somewhat lower than for similarly grown crystals where the carrier density corresponded to a Fermi level in the vicinity of $200$ meV.\cite{liangOng15,jeon14}

The Fermi level is about half of the interband transition onset energy,  indicating that the Dirac point is about midway between the final state (conduction band) and initial state (valence band), and the valence and conduction bands are more or less symmetrical. Band structure calculations and surface probe measurements show that the valence band is notably heavier than the conduction band, but that the two bands are not strongly asymmetrical.\cite{wang13,borisenkoCava14,jeon14,neupane14}

A very weak feature present at $\sim 2900 \text{ cm}^{-1}=360$ meV does not discernably shift with temperature (see in Fig. \ref{fig:CdAs}(c)) and is too high in energy to be associated with the Lifshitz gap energy. No optical signature of the Lifshitz gap is observed over the measured spectral region. However, even if it were within the measured range, it may not be optically measurable. The transition matrix elements in the vicinity of the $\Gamma$ point are expected to be suppressed like in the Na$_3$Bi case since the Dirac band orbital-characters are very similar.\cite{wang13}

\section{Conclusion}\label{sec:conclusion}
In both Cd$_3$As$_2$ and Na$_3$Bi, thermal occupation effects in the Dirac cone pair play a crucial  role in the optical response. Thermal excitation of carriers change the chemical potential and therefore the Dirac interband transition energy as well as the free carrier response.

In Cd$_3$As$_2$, the sharp Pauli-blocked edge at the onset of Dirac cone interband transitions induces a peak in the reflectivity with a very distinctive  lineshape, providing a fingerprint of the underlying Dirac cone dispersion and the associated logarithmic divergence in $\epsilon_1$. The frequency of the Pauli-blocked edge is controlled by the chemical potential that depends only on the power law exponent of the dispersion and the Fermi level in the low temperature limit. Our characterization of the peak location with temperature indicates a linear Dirac cone dispersion, a number density of $n=1.3 \times 10^{17}\text{ cm}^{-3}$, and a Fermi energy much larger than the Lifshitz gap energy as measured by STM.\cite{jeon14} The low temperature spectral width of the peak is caused by Fermi velocity anisotropy that gives rise to a narrow spectral range of Dirac cone interband transition onsets. The spectral width of the reflection peak translates into a Fermi velocity anisotropy of $10\%$, indicating a nearly spherical Fermi surface. The lineshape is incompatible with large Gaussian broadening effects, giving an upper bound energy scale for potential fluctuations of $\Gamma= 7$ meV.

In Na$_3$Bi, evidence of the Dirac cone manifests in a temperature dependent plasma edge caused by changes in the free carrier response. The Drude weight temperature dependence is nonmonotonic, attaining a minimum when the chemical potential is within $\sim kT$ of the Dirac node. The minimum in the temperature dependence of the plasma edge frequency at $T=100$K is characterized only by the Fermi level for a Dirac cone, giving $E_F=25$ meV. Unlike Cd$_3$As$_2$, evidence of the Dirac cone in Na$_3$Bi is not observable from the onset of Dirac cone interband transitions. The unobservable edge  presumably reflects the large Dirac cone anisotropy, which is consistent with band structure calculations. At transition energies well above the Lifshitz gap where the low energy Dirac cone pair has merged into one large Dirac cone, a frequency and temperature independent $\epsilon_2$ is observed. The constant value of $\epsilon_2$ is a fingerprint of the Dirac dispersion that only depends on Fermi velocity, and translates into an ab-plane Fermi velocity of $v_{\perp} \approx 3 \text{ eV} \AA$. The ground state of Na$_3$Bi has been reported as belonging to the $\text{P}6_3\text{/mmc}$ space group symmetry, but the number of observable IR active phonons that we observe rules this out in favor of the $\text{P}\bar{3}\text{c}1$ candidate symmetry. Finally, we have observed a plasmaron excitation near the plasma edge in Na$_3$Bi, which tracks the shifting ab-plane plasmon energy over a broad range of temperatures.

\section{Acknowledgments}
The work at UMD was supported by DOE under grant No. ER 46741-SC0005436. The research at Princeton was supported by the ARO MURI on topological insulators, Grant No. W911NF-12-1-0461 and ARO Grant No. W911NF-11-1-0379 and the MRSEC program at the Princeton Center for Complex Materials, Grant No. NSF-DMR-0819860 and Grant No. DOE DE-FG-02-05ER46200.
T.R.C. and H.T.J. are supported by the Ministry of Science and Technology,
National Tsing Hua University, and Academia Sinica, Taiwan, and they thank NCHC, CINC-NTU and NCTS, Taiwan for technical support. H.L. acknowledges the Singapore National Research Foundation for the support under NRF Award No. NRF-NRFF2013-03. The work at Northeastern University was supported by the US Department of Energy (DOE), Office of Science, Basic Energy Sciences grant number DE-FG02-07ER46352, and benefited from Northeastern University's Advanced Scientific Computation Center (ASCC) and the NERSC supercomputing center through DOE grant number DE-AC02-05CH11231. We thank Rolando V. Aguilar for useful conversations.

\appendix

\section{Chemical potential, Drude weight, and Pauli-blocked edge temperature dependence}\label{app:chempot}
The carrier density is given by $ n=\int_{-\infty}^\infty g(E) f(E) dE$, where $g(E)$ is the density of states and $f(E)$ is the Fermi distribution function. By fixing the number of carriers to the zero temperature value, the chemical potential $\mu$ for a given temperature $T$ is found by solving $\int_{-\infty}^{\infty} f(E) g(E) E dE = \int_{-\infty}^{E_F} g(E) E dE$.

An approximate expression for $\mu(T)$ is derived by the application of the Sommerfeld expansion assuming the temperature is much smaller than the Fermi energy $E_F$ and the integrand varies slowly over the energy range $E_F \pm T$, giving $\mu= E_F - \frac{1}{6} (\pi T)^2  \frac{g'(E_F)}{g(E_F)}+O(T^4)$, which is equation (2.77) in Ref. \onlinecite{ashcroft}. Expressing the density of states as $g = \frac{\partial n}{\partial k} /(\frac{\partial E}{\partial k})$, the carrier density as $n\propto k^3$, and a dispersion of the form $E \propto k^\beta$, leads to the expression $g'(E)/g(E)= E'(k)^{-1} g'(k)/g(k)= E^{-1}(3-\beta)/\beta$.

The DC conductivity $\sigma$ is derived in Chapter 13 of Ref. \onlinecite{ashcroft} and relates to the  Drude weight which becomes, after integration by parts, $D_W=\sigma/\tau \propto (1/v_F) \int E f(E) dE$ for a linear isotropic Dirac cone. Using the Sommerfeld expansion and substituting the expression for $\mu(T)$ gives $D_W(T)/D_W(0)= 1 -  \frac{1}{3}  (\frac{\pi T}{E_F})^2 +O(T^4)$, which agrees with results of Ref. \onlinecite{throckmortonPRB15}.

We now turn to derive the relationship between the chemical potential and the measured temperature-dependent frequency of the Pauli-blocked edge feature observed in Cd$_3$As$_2$. The optical Pauli-blocked edge frequency is given by $\omega(T)=\mu_{CB}(T)+E_{VB}(T)$ for a vertical (momentum conserving) transition between the final state in the conduction band at the chemical potential $\mu_{CB}(T)$ above the Dirac point, and an initial state at  $E_{VB}(T)$ in the valence band.  For Cd$_3$As$_2$, the conduction band chemical potential is much larger than the width of the Fermi distribution function for the temperature region of interest $T\leqslant150K$, so a negligible number of carriers will be thermally excited from the valence band. We approximate the conduction  band energy near $k=k_F$  by  $E \propto k^\beta$, which touches the valence band at a point. The valence band energy near $k=k_F$ is similarly approximated by $E\propto k^\beta$ that may have a different velocity from the conduction band. it is then straightforward to obtain the expression: $\omega(T)/\omega(0) = 1-\frac{1}{6} \frac{3-\beta}{\beta}(\frac{\pi T}{E_F})^2$.  The coefficient $C_{exp}$ is a fitting parameter found from the data $\delta \omega/\omega(0) = -C_{exp} T^2$. The Fermi energy is then calculated using $E_F=\sqrt{\frac{3-\beta}{6 \beta}\frac{\pi^2}{C_{exp}}}$, where $\beta=1$. Expanding $E_F$ about $\beta=1$ to first order gives $\frac{\delta E_F}{E_F|_{\beta=1}}\approx-\frac{3}{4 } \delta \beta$, so that if the dispersion tends toward superlinear near $k=k_F$, an estimation of $E_F$ using $\beta=1$ tends to overestimate $E_F$.

\section{Egg-shaped Fermi surface}\label{app:EggFS}
Consider an egg-shaped Fermi surface constructed with two half ellipsoids, each with a different major axis along $k_z$. The k-space volume is then given by $V_k = (4\pi/6) k_\perp^2 (k_{z1}+k_{z2})$. Assume a Dirac dispersion where $E=\hbar k_\perp v_\perp = \hbar k_{z1} v_{z1} =\hbar k_{z2} v_{z2}$ and $v_{z1}<<v_{z2},v_\perp$.  The carrier density is $n=N_d V_k/(2\pi)^3$, where $N_d=4$ is the degeneracy for a pair of Dirac cones. The applied electric field is assumed to be in the x-y plane, so the plasma frequency is given by $\Omega_P^2=4\pi n e^2 v_\perp^2/E_F$. Combining these results gives $E_F = \sqrt{3 \pi \hbar^3 v_{z1}/N_d} \Omega_P$. The temperature dependent chemical potential $\mu(T)/E_F$ and Drude weight $D_W(T)/D_W(0)$ under these assumed anisotropic velocity conditions is exactly the same as the isotropic Fermi velocity case (with $\beta=1$) derived above in Appendix \ref{app:chempot}.

The Pauli-blocked edge peak-feature shown in Fig. \ref{fig:ExpectedOpticalSigs} will be broadened by $\Delta \omega_{onset}$ via the velocity anisotropy $\alpha = v_{z1}/v_\perp$, where $v_{z2}$ is taken to equal $v_\perp$ for convenience. Here we derive the relationship between $\Delta \omega_{onset}$ and $\alpha$. For a Fermi energy lying in the conduction band, $E_F = \hbar k_\perp v_\perp=\hbar k_{z1} v_{z1}$ is the final state energy for Dirac cone interband transitions, and the extremum of initial state energies in the valence band is given by $E_{VB0}=\hbar k_\perp v_{z1}$ and $E_{VB1}=\hbar k_{z1} v_\perp$. Noting that $\Delta \omega=E_{VB1}-E_{VB0}$ and the average interband transition energy is $\bar{\omega}=E_F+(1/2)(E_{VB1}+E_{VB0})$ gives $\Delta \omega = E_F (1-\alpha^2)/\alpha$ and $\bar{\omega}= E_F (1+\alpha)^2/(2\alpha)$. Based on experimental data for Cd$_3$As$_2$, we obtain $\Delta \omega/\bar{\omega}=15/204$, and $\alpha=0.9$ .

\section{Phonon point group analysis}\label{app:PGAnalysis}
For Na$_3$Bi, the symmetries $P\bar{3}c1$, $P6_{3}/mmc$, and $P6_{3}cm$ have ground state energies that only differ by a few meV based on numerical calculations.\cite{cheng14} Phonon analysis of these three possible symmetries  are summarized in Tables \ref{tab:1}, \ref{tab:2}, and \ref{tab:3}.

\begin{table}
\caption{\label{tab:1} Phonon analysis for space group $P\bar{3}c1$ (\#165) with six Na$_3$Bi formula units per primitive cell.}
\mbox{
\begin{ruledtabular}
\begin{tabular}{cccc}
\multicolumn{4}{c}{Species$|$Wyckoff posn.$|$Site symm.$|$Vibrational modes}\\
\hline
Bi & 6f &  $C_2$ & $A_{1g}+A_{1u}+2A_{2g}+2A_{2u}+3E_{g}+3E_{u}$ \\

Na1 & 2a  & $D_3$ & $A_{2g} + A_{2u} + E_g + E_{u}$ \\

Na2 & 4d  & $C_3$ & $A_{1g} + A_{1u} + A_{2g} + A_{2u} +2E_{g}+2E_{u}$ \\

Na3 & 12g  & $C_1$ &  $3A_{1g} +3A_{1u} + 3A_{2g} + 3A_{2u} + 6E_{g}+ 6E_{u}$ \\

\hline
\multicolumn{2}{r}{Total:} & \multicolumn{2}{c}{$5A_{1g} + 5A_{1u} + 7A_{2g} + 7A_{2u} + 12E_g + 12E_{u}$} \\
\hline
\multicolumn{2}{r}{Acoustic:} & \multicolumn{2}{c}{$A_{2u} + E_{u}$} \\
\multicolumn{2}{r}{Infrared:} & \multicolumn{2}{c}{$6A_{2u}(e\parallel c) + 11E_{u}(e\perp c)$} \\
\multicolumn{2}{r}{Raman:} & \multicolumn{2}{c}{$5 A_{1g} + 12E_g $}\\
\multicolumn{2}{r}{Silent:} & \multicolumn{2}{c}{$7A_{2g} + 5A_{1u} + 12E_u$} \\
\end{tabular}
\end{ruledtabular}
}

\end{table}

\begin{table}
\caption{\label{tab:2} Phonon analysis for space group $P6_{3}cm$ (\#185) with six Na$_3$Bi formula units per primitive cell. }
\mbox{
\begin{ruledtabular}
\begin{tabular}{cccc}
\multicolumn{4}{c}{Species$|$Wyckoff posn.$|$Site symm.$|$Vibrational modes}\\
\hline
Bi & 6c &  $C_s$ & $2A_{1}+A_{2}+B_{1}+2B_{2}+3E_{1}+3E_{2}$ \\

Na1 & 2a  & $C_{3v}$ & $A_{1} + B_{2} + E_1 + E_{2}$ \\

Na2 & 4b  & $C_3$ & $A_{1} + A_{2} + B_{1} + B_{2} +2E_{1}+2E_{2}$ \\

Na3 & 6c &  $C_s$ & $2A_{1}+A_{2}+B_{1}+2B_{2}+3E_{1}+3E_{2}$ \\

Na4 & 6c &  $C_s$ & $2A_{1}+A_{2}+B_{1}+2B_{2}+3E_{1}+3E_{2}$ \\

\hline
\multicolumn{2}{r}{Total:} & \multicolumn{2}{c}{$8A_{1} + 4A_{2} + 4B_{1} + 8B_{2} + 12E_1 + 12E_{2}$} \\
\hline
\multicolumn{2}{r}{Acoustic:} & \multicolumn{2}{c}{$A_{1} + E_{1}$} \\
\multicolumn{2}{r}{Infrared:} & \multicolumn{2}{c}{$7A_{1}(e\parallel c) + 11E_{1}(e\perp c)$} \\
\multicolumn{2}{r}{Raman:} & \multicolumn{2}{c}{$7A_{1} + 11E_1+12E_2 $}\\
\multicolumn{2}{r}{Silent:} & \multicolumn{2}{c}{$4A_2+4B_1+8B_2$} \\
\end{tabular}
\end{ruledtabular}
}
\end{table}

\begin{table}
\caption{\label{tab:3} Phonon analysis for space group $P6_{3}/mmc$ (\#194) with two Na$_3$Bi formula units per primitive cell.}

\mbox{
\begin{ruledtabular}
\begin{tabular}{cccc}
\multicolumn{4}{c}{Species$|$Wyckoff posn.$|$Site symm.$|$Vibrational modes}\\
\hline
Bi & 2c &  $D_{3h}$ & $A_{2u}+B_{1g}+E_{1u}+E_{2g}$ \\

Na1 & 2b  & $D_{3h}$ & $A_{2u}+B_{1g}+E_{1u}+E_{2g}$ \\

Na2 & 4f  & $C_3$ & $A_{1g} + A_{2u} + B_{1g} + B_{2u} +E_{1g}$\\
\multicolumn{3}{c}{} & \multicolumn{1}{c}{$+E_{1u}+E_{2g}+E_{2u}$} \\

\hline
\multicolumn{2}{r}{Total:} & \multicolumn{2}{c}{$A_{1g} + 3A_{2u} + 3B_{1g} + B_{2u} + E_{1g} + E_{2g}$} \\
\multicolumn{2}{r}{} & \multicolumn{2}{c}{$+3E_{1u} + 3E_{2u}$} \\
\hline
\multicolumn{2}{r}{Acoustic:} & \multicolumn{2}{c}{$A_{2u}+E_{1u}$} \\
\multicolumn{2}{r}{Infrared:} & \multicolumn{2}{c}{$2A_{2u}(e\parallel c) + 2E_{1u}(e\perp c)$} \\
\multicolumn{2}{r}{Raman:} & \multicolumn{2}{c}{$A_{1g} + E_{1g}+E_{2g} $}\\
\multicolumn{2}{r}{Silent:} & \multicolumn{2}{c}{$3B_{1g}+B_{2u}+3E_{2u}$} \\

\end{tabular}
\end{ruledtabular}
}
\end{table}

\section{Estimate of $\epsilon_2$ from band structure calculations of the Dirac cone bands in the low frequency limit}\label{app:kdotp}
Here we obtain an estimate of $\epsilon_2$ in the low frequency limit based on $k\cdot p$ theory with fitting parameters that approximate the first-principles Dirac dispersion. The starting point is the formalism developed by Wang \emph{et al.}\cite{wangPRB12} in which the Dirac bands are described by a $4\times4$ leading order Hamiltonian around the $\Gamma$ point. Performing an expansion about the Dirac node such that $k_z \equiv k'_z -k_d$ where $2 k_d$ is the distance between Dirac nodes, and keeping up to linear terms in $k$ gives the following Hamiltonian:
\begin{align*}
& {\cal H}=2 C_1 \sqrt{\frac{M_0}{M_1}} k_z + \\
&                                          \begin{pmatrix}
                                              -2 \sqrt{M_0 M_1} k_z & A k_+ & 0 & 0 \\
                                              A k_- & 2 \sqrt{M_0 M_1} k_z & 0 & 0 \\
                                              0 & 0 & -2 \sqrt{M_0 M_1} k_z & -A k_-  \\
                                              0 & 0 & -A k_+ & 2 \sqrt{M_0 M_1} k_z \\
                                            \end{pmatrix}
\end{align*}
where $M_0, M_1, C_1$ and $A$ are parameters defined in Ref. \onlinecite{wang13} based on first-principles, and $k_\pm = k_x \pm i k_y$. The two $2\times2$ diagonal blocks in the Hamiltonian give the same eigenvalue solutions to leading order in $k$:  $E= v_0 k_z\pm \sqrt{(v_D k_z)^2 + (v_\perp k_\perp)^2}$, where $v_D^2=4 M_0 M_1, v_\perp=A, v_0 = 2 C_1 \sqrt{\frac{M_0}{M_1}}$, and $k_\perp^2=k_x^2 + k_y^2$. At $k_\perp=0$, the dispersion along $k_z$  gives slow and fast velocity solutions, $v_0\pm v_D$, near the Dirac node. ARPES measurements and  band structure calculations show that the velocity associated with the heavy Bi-like Dirac band ($v_{z1}$) is much smaller than the high velocity associated with the lighter Na-like band ($v_{z2}$), so that $v_0 \sim v_D$, and therefore, $v_D \approx v_{z2}/2$.

In the presence of an oscillating electric field in the x-y plane, the $4\times4$ interaction Hamiltonian contains two $2\times2$ diagonal blocks:
\begin{equation*}
{\cal H}_{int}=\pm\begin{pmatrix}
           0 & A A_\pm \\
           A A_\mp & 0 \\
         \end{pmatrix}
\end{equation*}
where $A_\pm= e \epsilon_\pm /(i\omega)$ is the vector potential and $\epsilon$ is the electric field, and the upper (lower) sign applies to the upper (lower) block. The square of the expectation value of the dipole matrix elements is given by $(e v_\perp \epsilon_\perp/\omega)^2$ and the joint density of states is given by $\frac{N_d}{6 \pi^2} \frac{(\hbar \omega)^3}{(2\hbar)^3 v_\perp^2 v_D}$. Using the Fermi's golden rule, the optical response is then simply obtained as:\cite{YuCardona}  $\epsilon_2=\frac{1}{6} N_d \alpha'$, where $\alpha'=e^2/\hbar v_D$. $\epsilon_2$ is seen to be independent of the transverse Fermi velocity $v_\perp$, being determined solely by the fast-velocity root of the z-component dispersion for the case of Na$_3$Bi where $v_D \approx v_{z2}/2$.

\bibliography{CdAsBib}

\begin{thebibliography}{10}
\expandafter\ifx\csname url\endcsname\relax
  \def\url#1{\texttt{#1}}\fi
\expandafter\ifx\csname urlprefix\endcsname\relax\def\urlprefix{URL }\fi
\providecommand{\bibinfo}[2]{#2}
\providecommand{\eprint}[2][]{\url{#2}}

\bibitem{bansilrev16}
\bibinfo{author}{Bansil, A.}, \bibinfo{author}{Lin, H.} \&
  \bibinfo{author}{Das, T.}
\newblock \bibinfo{title}{Colloquium: {Topological} {Band} {Theory}}.
\newblock \emph{\bibinfo{journal}{arXiv:1603.03576 [cond-mat]}}
  (\bibinfo{year}{2016}).
\newblock \urlprefix\url{http://arxiv.org/abs/1603.03576}.
\newblock \bibinfo{note}{ArXiv: 1603.03576}.

\bibitem{kane10}
\bibinfo{author}{Hasan, M.~Z.} \& \bibinfo{author}{Kane, C.~L.}
\newblock \bibinfo{title}{Colloquium: {Topological} insulators}.
\newblock \emph{\bibinfo{journal}{Reviews of Modern Physics}}
  \textbf{\bibinfo{volume}{82}}, \bibinfo{pages}{3045} (\bibinfo{year}{2010}).
\newblock \urlprefix\url{http://link.aps.org/doi/10.1103/RevModPhys.82.3045}.

\bibitem{qi11}
\bibinfo{author}{Qi, X.-L.} \& \bibinfo{author}{Zhang, S.-C.}
\newblock \bibinfo{title}{Topological insulators and superconductors}.
\newblock \emph{\bibinfo{journal}{Reviews of Modern Physics}}
  \textbf{\bibinfo{volume}{83}}, \bibinfo{pages}{1057--1110}
  (\bibinfo{year}{2011}).
\newblock \urlprefix\url{http://link.aps.org/doi/10.1103/RevModPhys.83.1057}.

\bibitem{Moore11}
\bibinfo{author}{Hasan, M.~Z.} \& \bibinfo{author}{Moore, J.~E.}
\newblock \bibinfo{title}{Three-{Dimensional} {Topological} {Insulators}}.
\newblock \emph{\bibinfo{journal}{Annual Review of Condensed Matter Physics}}
  \textbf{\bibinfo{volume}{2}}, \bibinfo{pages}{55--78} (\bibinfo{year}{2011}).
\newblock
  \urlprefix\url{http://dx.doi.org/10.1146/annurev-conmatphys-062910-140432}.

\bibitem{wang12}
\bibinfo{author}{Wang, Z.} \emph{et~al.}
\newblock \bibinfo{title}{Dirac semimetal and topological phase transitions in
  {A}3bi ({A}={Na}, {K}, {Rb})}.
\newblock \emph{\bibinfo{journal}{Physical Review B}}
  \textbf{\bibinfo{volume}{85}}, \bibinfo{pages}{195320}
  (\bibinfo{year}{2012}).
\newblock \urlprefix\url{http://link.aps.org/doi/10.1103/PhysRevB.85.195320}.

\bibitem{wang13}
\bibinfo{author}{Wang, Z.}, \bibinfo{author}{Weng, H.}, \bibinfo{author}{Wu,
  Q.}, \bibinfo{author}{Dai, X.} \& \bibinfo{author}{Fang, Z.}
\newblock \bibinfo{title}{Three-dimensional {Dirac} semimetal and quantum
  transport in {Cd}\$\{\}\_\{3\}\${As}\$\{\}\_\{2\}\$}.
\newblock \emph{\bibinfo{journal}{Physical Review B}}
  \textbf{\bibinfo{volume}{88}}, \bibinfo{pages}{125427}
  (\bibinfo{year}{2013}).
\newblock \urlprefix\url{http://link.aps.org/doi/10.1103/PhysRevB.88.125427}.

\bibitem{xuHasan13}
\bibinfo{author}{Xu, S.-Y.} \emph{et~al.}
\newblock \bibinfo{title}{Observation of {Fermi} arc surface states in a
  topological metal}.
\newblock \emph{\bibinfo{journal}{Science}} \textbf{\bibinfo{volume}{347}},
  \bibinfo{pages}{294--298} (\bibinfo{year}{2015}).
\newblock \urlprefix\url{http://www.sciencemag.org/content/347/6219/294}.

\bibitem{LiuZXShen14}
\bibinfo{author}{Liu, Z.~K.} \emph{et~al.}
\newblock \bibinfo{title}{Discovery of a {Three}-{Dimensional} {Topological}
  {Dirac} {Semimetal}, {Na}3bi}.
\newblock \emph{\bibinfo{journal}{Science}} \textbf{\bibinfo{volume}{343}},
  \bibinfo{pages}{864--867} (\bibinfo{year}{2014}).
\newblock \urlprefix\url{http://www.sciencemag.org/content/343/6173/864}.

\bibitem{xiongOngCA15}
\bibinfo{author}{Xiong, J.} \emph{et~al.}
\newblock \bibinfo{title}{Signature of the chiral anomaly in a {Dirac}
  semimetal: a current plume steered by a magnetic field}.
\newblock \emph{\bibinfo{journal}{arXiv:1503.08179 [cond-mat]}}
  (\bibinfo{year}{2015}).
\newblock \urlprefix\url{http://arxiv.org/abs/1503.08179}.
\newblock \bibinfo{note}{ArXiv: 1503.08179}.

\bibitem{neupane14}
\bibinfo{author}{Neupane, M.} \emph{et~al.}
\newblock \bibinfo{title}{Observation of a three-dimensional topological
  {Dirac} semimetal phase in high-mobility {Cd}3as2}.
\newblock \emph{\bibinfo{journal}{Nature Communications}}
  \textbf{\bibinfo{volume}{5}} (\bibinfo{year}{2014}).
\newblock
  \urlprefix\url{http://www.nature.com/ncomms/2014/140507/ncomms4786/full/ncomms4786.html}.

\bibitem{liuNature14}
\bibinfo{author}{Liu, Z.~K.} \emph{et~al.}
\newblock \bibinfo{title}{A stable three-dimensional topological {Dirac}
  semimetal {Cd}3as2}.
\newblock \emph{\bibinfo{journal}{Nature Materials}}
  \textbf{\bibinfo{volume}{13}}, \bibinfo{pages}{677--681}
  (\bibinfo{year}{2014}).
\newblock
  \urlprefix\url{http://www.nature.com/nmat/journal/v13/n7/full/nmat3990.html}.

\bibitem{jeon14}
\bibinfo{author}{Jeon, S.} \emph{et~al.}
\newblock \bibinfo{title}{Landau quantization and quasiparticle interference in
  the three-dimensional {Dirac} semimetal {Cd}3as2}.
\newblock \emph{\bibinfo{journal}{Nature Materials}}
  \textbf{\bibinfo{volume}{13}}, \bibinfo{pages}{851--856}
  (\bibinfo{year}{2014}).
\newblock
  \urlprefix\url{http://www.nature.com/nmat/journal/v13/n9/full/nmat4023.html}.

\bibitem{yi14}
\bibinfo{author}{Yi, H.} \emph{et~al.}
\newblock \bibinfo{title}{Evidence of {Topological} {Surface} {State} in
  {Three}-{Dimensional} {Dirac} {Semimetal} {Cd}3as2}.
\newblock \emph{\bibinfo{journal}{Scientific Reports}}
  \textbf{\bibinfo{volume}{4}} (\bibinfo{year}{2014}).
\newblock
  \urlprefix\url{http://www.nature.com/srep/2014/140820/srep06106/full/srep06106.html}.

\bibitem{borisenkoCava14}
\bibinfo{author}{Borisenko, S.} \emph{et~al.}
\newblock \bibinfo{title}{Experimental {Realization} of a {Three}-{Dimensional}
  {Dirac} {Semimetal}}.
\newblock \emph{\bibinfo{journal}{Physical Review Letters}}
  \textbf{\bibinfo{volume}{113}}, \bibinfo{pages}{027603}
  (\bibinfo{year}{2014}).
\newblock
  \urlprefix\url{http://link.aps.org/doi/10.1103/PhysRevLett.113.027603}.

\bibitem{wan2011}
\bibinfo{author}{Wan, X.}, \bibinfo{author}{Turner, A.~M.},
  \bibinfo{author}{Vishwanath, A.} \& \bibinfo{author}{Savrasov, S.~Y.}
\newblock \bibinfo{title}{Topological semimetal and {Fermi}-arc surface states
  in the electronic structure of pyrochlore iridates}.
\newblock \emph{\bibinfo{journal}{Physical Review B}}
  \textbf{\bibinfo{volume}{83}}, \bibinfo{pages}{205101}
  (\bibinfo{year}{2011}).
\newblock \urlprefix\url{http://link.aps.org/doi/10.1103/PhysRevB.83.205101}.

\bibitem{burkov11}
\bibinfo{author}{Burkov, A.~A.} \& \bibinfo{author}{Balents, L.}
\newblock \bibinfo{title}{Weyl {Semimetal} in a {Topological} {Insulator}
  {Multilayer}}.
\newblock \emph{\bibinfo{journal}{Physical Review Letters}}
  \textbf{\bibinfo{volume}{107}}, \bibinfo{pages}{127205}
  (\bibinfo{year}{2011}).
\newblock
  \urlprefix\url{http://link.aps.org/doi/10.1103/PhysRevLett.107.127205}.

\bibitem{weng14}
\bibinfo{author}{Weng, H.}, \bibinfo{author}{Fang, C.}, \bibinfo{author}{Fang,
  Z.}, \bibinfo{author}{Bernevig, B.~A.} \& \bibinfo{author}{Dai, X.}
\newblock \bibinfo{title}{Weyl semimetal phase in noncentrosymmetric
  transition-metal monophosphides}.
\newblock \emph{\bibinfo{journal}{Physical Review X}}
  \textbf{\bibinfo{volume}{5}}, \bibinfo{pages}{011029} (\bibinfo{year}{2015}).
\newblock \urlprefix\url{http://link.aps.org/doi/10.1103/PhysRevX.5.011029}.

\bibitem{huang15}
\bibinfo{author}{Huang, S.-M.} \emph{et~al.}
\newblock \bibinfo{title}{A weyl fermion semimetal with surface fermi arcs in
  the transition metal monopnictide {TaAs} class}.
\newblock \emph{\bibinfo{journal}{Nature Communications}}
  \textbf{\bibinfo{volume}{6}}, \bibinfo{pages}{7373} (\bibinfo{year}{2015}).
\newblock
  \urlprefix\url{http://www.nature.com/ncomms/2015/150612/ncomms8373/full/ncomms8373.html}.

\bibitem{lv15}
\bibinfo{author}{Lv, B.} \emph{et~al.}
\newblock \bibinfo{title}{Experimental discovery of weyl semimetal {TaAs}}.
\newblock \emph{\bibinfo{journal}{Physical Review X}}
  \textbf{\bibinfo{volume}{5}}, \bibinfo{pages}{031013} (\bibinfo{year}{2015}).
\newblock \urlprefix\url{http://link.aps.org/doi/10.1103/PhysRevX.5.031013}.

\bibitem{xu15}
\bibinfo{author}{Xu, S.-Y.} \emph{et~al.}
\newblock \bibinfo{title}{Discovery of a weyl fermion semimetal and topological
  fermi arcs}.
\newblock \emph{\bibinfo{journal}{Science}} \textbf{\bibinfo{volume}{349}},
  \bibinfo{pages}{613--617} (\bibinfo{year}{2015}).
\newblock \urlprefix\url{http://www.sciencemag.org/content/349/6248/613}.

\bibitem{lvn215}
\bibinfo{author}{Lv, B.~Q.} \emph{et~al.}
\newblock \bibinfo{title}{Observation of weyl nodes in {TaAs}}.
\newblock \emph{\bibinfo{journal}{Nature Physics}}
  \textbf{\bibinfo{volume}{11}}, \bibinfo{pages}{724--727}
  (\bibinfo{year}{2015}).
\newblock
  \urlprefix\url{http://www.nature.com/nphys/journal/v11/n9/full/nphys3426.html}.

\bibitem{fu08}
\bibinfo{author}{Fu, L.} \& \bibinfo{author}{Kane, C.~L.}
\newblock \bibinfo{title}{Superconducting {Proximity} {Effect} and {Majorana}
  {Fermions} at the {Surface} of a {Topological} {Insulator}}.
\newblock \emph{\bibinfo{journal}{Physical Review Letters}}
  \textbf{\bibinfo{volume}{100}}, \bibinfo{pages}{096407}
  (\bibinfo{year}{2008}).
\newblock
  \urlprefix\url{http://link.aps.org/doi/10.1103/PhysRevLett.100.096407}.

\bibitem{mourik12}
\bibinfo{author}{Mourik, V.} \emph{et~al.}
\newblock \bibinfo{title}{Signatures of {Majorana} {Fermions} in {Hybrid}
  {Superconductor}-{Semiconductor} {Nanowire} {Devices}}.
\newblock \emph{\bibinfo{journal}{Science}} \textbf{\bibinfo{volume}{336}},
  \bibinfo{pages}{1003--1007} (\bibinfo{year}{2012}).
\newblock \urlprefix\url{http://www.sciencemag.org/content/336/6084/1003}.

\bibitem{nadj14}
\bibinfo{author}{Nadj-Perge, S.} \emph{et~al.}
\newblock \bibinfo{title}{Observation of {Majorana} fermions in ferromagnetic
  atomic chains on a superconductor}.
\newblock \emph{\bibinfo{journal}{Science}} \textbf{\bibinfo{volume}{346}},
  \bibinfo{pages}{602--607} (\bibinfo{year}{2014}).
\newblock \urlprefix\url{http://www.sciencemag.org/content/346/6209/602}.

\bibitem{zhong15}
\bibinfo{author}{Zhong, S.}, \bibinfo{author}{Orenstein, J.} \&
  \bibinfo{author}{Moore, J.~E.}
\newblock \bibinfo{title}{Optical {Gyrotropy} from {Axion} {Electrodynamics} in
  {Momentum} {Space}}.
\newblock \emph{\bibinfo{journal}{Physical Review Letters}}
  \textbf{\bibinfo{volume}{115}}, \bibinfo{pages}{117403}
  (\bibinfo{year}{2015}).
\newblock
  \urlprefix\url{http://link.aps.org/doi/10.1103/PhysRevLett.115.117403}.

\bibitem{goswami13}
\bibinfo{author}{Goswami, P.} \& \bibinfo{author}{Tewari, S.}
\newblock \bibinfo{title}{Axionic field theory of \$(3+1)\$-dimensional {Weyl}
  semimetals}.
\newblock \emph{\bibinfo{journal}{Physical Review B}}
  \textbf{\bibinfo{volume}{88}}, \bibinfo{pages}{245107}
  (\bibinfo{year}{2013}).
\newblock \urlprefix\url{http://link.aps.org/doi/10.1103/PhysRevB.88.245107}.

\bibitem{zyuzin12}
\bibinfo{author}{Zyuzin, A.~A.} \& \bibinfo{author}{Burkov, A.~A.}
\newblock \bibinfo{title}{Topological response in {Weyl} semimetals and the
  chiral anomaly}.
\newblock \emph{\bibinfo{journal}{Physical Review B}}
  \textbf{\bibinfo{volume}{86}}, \bibinfo{pages}{115133}
  (\bibinfo{year}{2012}).
\newblock \urlprefix\url{http://link.aps.org/doi/10.1103/PhysRevB.86.115133}.

\bibitem{fujikawa79}
\bibinfo{author}{Fujikawa, K.}
\newblock \bibinfo{title}{Path-{Integral} {Measure} for {Gauge}-{Invariant}
  {Fermion} {Theories}}.
\newblock \emph{\bibinfo{journal}{Physical Review Letters}}
  \textbf{\bibinfo{volume}{42}}, \bibinfo{pages}{1195--1198}
  (\bibinfo{year}{1979}).
\newblock \urlprefix\url{http://link.aps.org/doi/10.1103/PhysRevLett.42.1195}.

\bibitem{xiongOng15}
\bibinfo{author}{Xiong, J.} \emph{et~al.}
\newblock \bibinfo{title}{Anomalous conductivity tensor in the dirac semimetal
  na 3 bi}.
\newblock \emph{\bibinfo{journal}{{EPL} (Europhysics Letters)}}
  \textbf{\bibinfo{volume}{114}}, \bibinfo{pages}{27002}
  (\bibinfo{year}{2016}).
\newblock \urlprefix\url{http://stacks.iop.org/0295-5075/114/i=2/a=27002}.

\bibitem{Hofmann14}
\bibinfo{author}{Hofmann, J.}, \bibinfo{author}{Barnes, E.} \&
  \bibinfo{author}{Das~Sarma, S.}
\newblock \bibinfo{title}{Interacting dirac liquid in three-dimensional
  semimetals}.
\newblock \emph{\bibinfo{journal}{Physical Review B}}
  \textbf{\bibinfo{volume}{92}}, \bibinfo{pages}{045104}
  (\bibinfo{year}{2015}).
\newblock \urlprefix\url{http://link.aps.org/doi/10.1103/PhysRevB.92.045104}.

\bibitem{Hofmann15}
\bibinfo{author}{Hofmann, J.} \& \bibinfo{author}{Das~Sarma, S.}
\newblock \bibinfo{title}{Plasmon signature in dirac-weyl liquids}.
\newblock \emph{\bibinfo{journal}{Physical Review B}}
  \textbf{\bibinfo{volume}{91}}, \bibinfo{pages}{241108}
  (\bibinfo{year}{2015}).
\newblock \urlprefix\url{http://link.aps.org/doi/10.1103/PhysRevB.91.241108}.

\bibitem{ashby14}
\bibinfo{author}{Ashby, P. E.~C.} \& \bibinfo{author}{Carbotte, J.~P.}
\newblock \bibinfo{title}{Chiral anomaly and optical absorption in {Weyl}
  semimetals}.
\newblock \emph{\bibinfo{journal}{Physical Review B}}
  \textbf{\bibinfo{volume}{89}}, \bibinfo{pages}{245121}
  (\bibinfo{year}{2014}).
\newblock \urlprefix\url{http://link.aps.org/doi/10.1103/PhysRevB.89.245121}.

\bibitem{hosur12}
\bibinfo{author}{Hosur, P.}, \bibinfo{author}{Parameswaran, S.} \&
  \bibinfo{author}{Vishwanath, A.}
\newblock \bibinfo{title}{Charge {Transport} in {Weyl} {Semimetals}}.
\newblock \emph{\bibinfo{journal}{Physical Review Letters}}
  \textbf{\bibinfo{volume}{108}}, \bibinfo{pages}{046602}
  (\bibinfo{year}{2012}).
\newblock
  \urlprefix\url{http://link.aps.org/doi/10.1103/PhysRevLett.108.046602}.

\bibitem{skinner14}
\bibinfo{author}{Skinner, B.}
\newblock \bibinfo{title}{Coulomb disorder in three-dimensional {Dirac}
  systems}.
\newblock \emph{\bibinfo{journal}{Physical Review B}}
  \textbf{\bibinfo{volume}{90}}, \bibinfo{pages}{060202}
  (\bibinfo{year}{2014}).
\newblock \urlprefix\url{http://link.aps.org/doi/10.1103/PhysRevB.90.060202}.

\bibitem{throckmortonPRB15}
\bibinfo{author}{Throckmorton, R.~E.}, \bibinfo{author}{Hofmann, J.},
  \bibinfo{author}{Barnes, E.} \& \bibinfo{author}{Das~Sarma, S.}
\newblock \bibinfo{title}{Many-body effects and ultraviolet renormalization in
  three-dimensional {Dirac} materials}.
\newblock \emph{\bibinfo{journal}{Physical Review B}}
  \textbf{\bibinfo{volume}{92}}, \bibinfo{pages}{115101}
  (\bibinfo{year}{2015}).
\newblock \urlprefix\url{http://link.aps.org/doi/10.1103/PhysRevB.92.115101}.

\bibitem{tediosi2007}
\bibinfo{author}{Tediosi, R.}, \bibinfo{author}{Armitage, N.~P.},
  \bibinfo{author}{Giannini, E.} \& \bibinfo{author}{van~der Marel, D.}
\newblock \bibinfo{title}{Charge {Carrier} {Interaction} with a {Purely}
  {Electronic} {Collective} {Mode}: {Plasmarons} and the {Infrared} {Response}
  of {Elemental} {Bismuth}}.
\newblock \emph{\bibinfo{journal}{Physical Review Letters}}
  \textbf{\bibinfo{volume}{99}}, \bibinfo{pages}{016406}
  (\bibinfo{year}{2007}).
\newblock
  \urlprefix\url{http://link.aps.org/doi/10.1103/PhysRevLett.99.016406}.

\bibitem{sonspivak13}
\bibinfo{author}{Son, D.~T.} \& \bibinfo{author}{Spivak, B.~Z.}
\newblock \bibinfo{title}{Chiral anomaly and classical negative
  magnetoresistance of {Weyl} metals}.
\newblock \emph{\bibinfo{journal}{Physical Review B}}
  \textbf{\bibinfo{volume}{88}}, \bibinfo{pages}{104412}
  (\bibinfo{year}{2013}).
\newblock \urlprefix\url{http://link.aps.org/doi/10.1103/PhysRevB.88.104412}.

\bibitem{goswami15}
\bibinfo{author}{Goswami, P.}, \bibinfo{author}{Sharma, G.} \&
  \bibinfo{author}{Tewari, S.}
\newblock \bibinfo{title}{Optical activity as a test for dynamic chiral
  magnetic effect of weyl semimetals}.
\newblock \emph{\bibinfo{journal}{Physical Review B}}
  \textbf{\bibinfo{volume}{92}}, \bibinfo{pages}{161110}
  (\bibinfo{year}{2015}).
\newblock \urlprefix\url{http://link.aps.org/doi/10.1103/PhysRevB.92.161110}.

\bibitem{kargarian15}
\bibinfo{author}{Kargarian, M.}, \bibinfo{author}{Randeria, M.} \&
  \bibinfo{author}{Trivedi, N.}
\newblock \bibinfo{title}{Theory of kerr and faraday rotations and linear
  dichroism in topological weyl semimetals}.
\newblock \emph{\bibinfo{journal}{Scientific Reports}}
  \textbf{\bibinfo{volume}{5}}, \bibinfo{pages}{12683} (\bibinfo{year}{2015}).
\newblock \urlprefix\url{http://www.nature.com/doifinder/10.1038/srep12683}.

\bibitem{hofmann16}
\bibinfo{author}{Hofmann, J.} \& \bibinfo{author}{Das~Sarma, S.}
\newblock \bibinfo{title}{Surface plasmon polaritons in topological weyl
  semimetals}.
\newblock \emph{\bibinfo{journal}{Physical Review B}}
  \textbf{\bibinfo{volume}{93}}, \bibinfo{pages}{241402}
  (\bibinfo{year}{2016}).
\newblock \urlprefix\url{http://link.aps.org/doi/10.1103/PhysRevB.93.241402}.

\bibitem{kushwaha15}
\bibinfo{author}{Kushwaha, S.~K.} \emph{et~al.}
\newblock \bibinfo{title}{Bulk crystal growth and electronic characterization
  of the 3d {Dirac} semimetal {Na}3bi}.
\newblock \emph{\bibinfo{journal}{APL Materials}} \textbf{\bibinfo{volume}{3}},
  \bibinfo{pages}{041504} (\bibinfo{year}{2015}).
\newblock
  \urlprefix\url{http://scitation.aip.org/content/aip/journal/aplmater/3/4/10.1063/1.4908158}.

\bibitem{rosenberg59}
\bibinfo{author}{Rosenberg, A.~J.} \& \bibinfo{author}{Harman, T.~C.}
\newblock \bibinfo{title}{Cd3as2—{A} {Noncubic} {Semiconductor} with
  {Unusually} {High} {Electron} {Mobility}}.
\newblock \emph{\bibinfo{journal}{Journal of Applied Physics}}
  \textbf{\bibinfo{volume}{30}}, \bibinfo{pages}{1621--1622}
  (\bibinfo{year}{1959}).
\newblock
  \urlprefix\url{http://scitation.aip.org/content/aip/journal/jap/30/10/10.1063/1.1735019}.

\bibitem{liangOng15}
\bibinfo{author}{Liang, T.} \emph{et~al.}
\newblock \bibinfo{title}{Ultrahigh mobility and giant magnetoresistance in the
  {Dirac} semimetal {Cd}3as2}.
\newblock \emph{\bibinfo{journal}{Nature Materials}}
  \textbf{\bibinfo{volume}{14}}, \bibinfo{pages}{280--284}
  (\bibinfo{year}{2015}).
\newblock
  \urlprefix\url{http://www.nature.com/nmat/journal/v14/n3/full/nmat4143.html}.

\bibitem{arush92}
\bibinfo{author}{Arushanov, E.~K.}
\newblock \bibinfo{title}{{II}3v2 compounds and alloys}.
\newblock \emph{\bibinfo{journal}{Progress in Crystal Growth and
  Characterization of Materials}} \textbf{\bibinfo{volume}{25}},
  \bibinfo{pages}{131--201} (\bibinfo{year}{1992}).
\newblock
  \urlprefix\url{http://www.sciencedirect.com/science/article/pii/096089749290030T}.

\bibitem{turner1961}
\bibinfo{author}{Turner, W.~J.}, \bibinfo{author}{Fischler, A.~S.} \&
  \bibinfo{author}{Reese, W.~E.}
\newblock \bibinfo{title}{Electrical and {Optical} {Properties} of the
  {II}–{V} {Compounds}}.
\newblock \emph{\bibinfo{journal}{Journal of Applied Physics}}
  \textbf{\bibinfo{volume}{32}}, \bibinfo{pages}{2241--2245}
  (\bibinfo{year}{1961}).
\newblock
  \urlprefix\url{http://scitation.aip.org/content/aip/journal/jap/32/10/10.1063/1.1777051}.

\bibitem{haide66}
\bibinfo{author}{Haidemenakis, E.~D.}, \bibinfo{author}{Mavroides, J.~G.},
  \bibinfo{author}{Dresselhaus, M.~S.} \& \bibinfo{author}{Kolesar, D.~F.}
\newblock \bibinfo{title}{Observation of interband transitions in {Cd}3as2}.
\newblock \emph{\bibinfo{journal}{Solid State Communications}}
  \textbf{\bibinfo{volume}{4}}, \bibinfo{pages}{65--68} (\bibinfo{year}{1966}).
\newblock
  \urlprefix\url{http://www.sciencedirect.com/science/article/pii/0038109866901074}.

\bibitem{gelten80}
\bibinfo{author}{Gelten, M.~J.}, \bibinfo{author}{van Es, C.~M.},
  \bibinfo{author}{Blom, F. A.~P.} \& \bibinfo{author}{Jongeneelen, J. W.~F.}
\newblock \bibinfo{title}{Optical verification of the valence band structure of
  cadmium arsenide}.
\newblock \emph{\bibinfo{journal}{Solid State Communications}}
  \textbf{\bibinfo{volume}{33}}, \bibinfo{pages}{833--836}
  (\bibinfo{year}{1980}).
\newblock
  \urlprefix\url{http://www.sciencedirect.com/science/article/pii/0038109880911990}.

\bibitem{houde86}
\bibinfo{author}{Houde, D.}, \bibinfo{author}{Jandl, S.},
  \bibinfo{author}{Banville, M.} \& \bibinfo{author}{Aubin, M.}
\newblock \bibinfo{title}{The infrared spectrum of {Cd}3as2}.
\newblock \emph{\bibinfo{journal}{Solid State Communications}}
  \textbf{\bibinfo{volume}{57}}, \bibinfo{pages}{247--248}
  (\bibinfo{year}{1986}).
\newblock
  \urlprefix\url{http://www.sciencedirect.com/science/article/pii/0038109886901493}.

\bibitem{neubauer16}
\bibinfo{author}{Neubauer, D.} \emph{et~al.}
\newblock \bibinfo{title}{Interband optical conductivity of the [001]-oriented
  dirac semimetal
  \$\{{\textbackslash}mathrm\{Cd\}\}\_\{3\}\{{\textbackslash}mathrm\{As\}\}\_\{2\}\$}.
\newblock \emph{\bibinfo{journal}{Physical Review B}}
  \textbf{\bibinfo{volume}{93}}, \bibinfo{pages}{121202}
  (\bibinfo{year}{2016}).
\newblock \urlprefix\url{http://link.aps.org/doi/10.1103/PhysRevB.93.121202}.

\bibitem{akrap2016}
\bibinfo{author}{Akrap, A.} \emph{et~al.}
\newblock \bibinfo{title}{Magneto-optical signature of massless kane electrons
  in cd3as2}.
\newblock \emph{\bibinfo{journal}{{arXiv}:1604.00038 [cond-mat]}}
  (\bibinfo{year}{2016}).
\newblock \urlprefix\url{http://arxiv.org/abs/1604.00038}.
\newblock \eprint{1604.00038}.

\bibitem{wangPRB12}
\bibinfo{author}{Wang, Z.} \emph{et~al.}
\newblock \bibinfo{title}{Dirac semimetal and topological phase transitions in
  {A}3bi ({A}={Na}, {K}, {Rb})}.
\newblock \emph{\bibinfo{journal}{Physical Review B}}
  \textbf{\bibinfo{volume}{85}}, \bibinfo{pages}{195320}
  (\bibinfo{year}{2012}).
\newblock \urlprefix\url{http://link.aps.org/doi/10.1103/PhysRevB.85.195320}.

\bibitem{YuCardona}
\bibinfo{author}{Yu, P.~Y.} \& \bibinfo{author}{Cardona, M.}
\newblock \emph{\bibinfo{title}{Fundamentals of {Semiconductors}}}.
\newblock Graduate {Texts} in {Physics} (\bibinfo{publisher}{Springer Berlin
  Heidelberg}, \bibinfo{address}{Berlin, Heidelberg}, \bibinfo{year}{2010}).
\newblock \urlprefix\url{http://link.springer.com/10.1007/978-3-642-00710-1}.

\bibitem{narayan14}
\bibinfo{author}{Narayan, A.}, \bibinfo{author}{Di~Sante, D.},
  \bibinfo{author}{Picozzi, S.} \& \bibinfo{author}{Sanvito, S.}
\newblock \bibinfo{title}{Topological tuning in three-dimensional dirac
  semimetals}.
\newblock \emph{\bibinfo{journal}{Physical Review Letters}}
  \textbf{\bibinfo{volume}{113}}, \bibinfo{pages}{256403}
  (\bibinfo{year}{2014}).
\newblock
  \urlprefix\url{http://link.aps.org/doi/10.1103/PhysRevLett.113.256403}.

\bibitem{cheng14}
\bibinfo{author}{Cheng, X.} \emph{et~al.}
\newblock \bibinfo{title}{Ground-state phase in the three-dimensional
  topological {Dirac} semimetal
  \$\{{\textbackslash}mathrm\{{Na}\}\}\_\{3\}{\textbackslash}mathrm\{{Bi}\}\$}.
\newblock \emph{\bibinfo{journal}{Physical Review B}}
  \textbf{\bibinfo{volume}{89}}, \bibinfo{pages}{245201}
  (\bibinfo{year}{2014}).
\newblock \urlprefix\url{http://link.aps.org/doi/10.1103/PhysRevB.89.245201}.

\bibitem{sushkov2015}
\bibinfo{author}{Sushkov, A.~B.} \emph{et~al.}
\newblock \bibinfo{title}{Optical evidence for a weyl semimetal state in
  pyrochlore
  \$\{{\textbackslash}mathrm\{Eu\}\}\_\{2\}\{{\textbackslash}mathrm\{Ir\}\}\_\{2\}\{{\textbackslash}mathrm\{O\}\}\_\{7\}\$}.
\newblock \emph{\bibinfo{journal}{Physical Review B}}
  \textbf{\bibinfo{volume}{92}}, \bibinfo{pages}{241108}
  (\bibinfo{year}{2015}).
\newblock \urlprefix\url{http://link.aps.org/doi/10.1103/PhysRevB.92.241108}.

\bibitem{ashcroft}
\bibinfo{author}{Ashcroft, N.~W.} \& \bibinfo{author}{Mermin, N.~D.}
\newblock \emph{\bibinfo{title}{Solid State Physics}}
  (\bibinfo{publisher}{Holt, Rinehart and Winston}, \bibinfo{year}{1976}).

\bibitem{throckmorton15}
\bibinfo{author}{Throckmorton, R.~E.}, \bibinfo{author}{Hofmann, J.},
  \bibinfo{author}{Barnes, E.} \& \bibinfo{author}{Das~Sarma, S.}
\newblock \bibinfo{title}{Many-body effects and ultraviolet renormalization in
  three-dimensional dirac materials}.
\newblock \emph{\bibinfo{journal}{Physical Review B}}
  \textbf{\bibinfo{volume}{92}}, \bibinfo{pages}{115101}
  (\bibinfo{year}{2015}).
\newblock \urlprefix\url{http://link.aps.org/doi/10.1103/PhysRevB.92.115101}.

\bibitem{kuzmenkoVDF2005}
\bibinfo{author}{Kuzmenko, A.~B.}
\newblock \bibinfo{title}{Kramers–{Kronig} constrained variational analysis
  of optical spectra}.
\newblock \emph{\bibinfo{journal}{Review of Scientific Instruments}}
  \textbf{\bibinfo{volume}{76}}, \bibinfo{pages}{083108}
  (\bibinfo{year}{2005}).
\newblock
  \urlprefix\url{http://scitation.aip.org/content/aip/journal/rsi/76/8/10.1063/1.1979470}.

\bibitem{lundqvist67}
\bibinfo{author}{Lundqvist, B.~I.}
\newblock \bibinfo{title}{Single-particle spectrum of the degenerate electron
  gas}.
\newblock \emph{\bibinfo{journal}{Physik der kondensierten Materie}}
  \textbf{\bibinfo{volume}{6}}, \bibinfo{pages}{193--205}
  (\bibinfo{year}{1967}).
\newblock \urlprefix\url{http://link.springer.com/article/10.1007/BF02422716}.

\bibitem{armitageBi10}
\bibinfo{author}{Armitage, N.~P.} \emph{et~al.}
\newblock \bibinfo{title}{Infrared {Conductivity} of {Elemental} {Bismuth}
  under {Pressure}: {Evidence} for an {Avoided} {Lifshitz}-{Type}
  {Semimetal}-{Semiconductor} {Transition}}.
\newblock \emph{\bibinfo{journal}{Physical Review Letters}}
  \textbf{\bibinfo{volume}{104}}, \bibinfo{pages}{237401}
  (\bibinfo{year}{2010}).
\newblock
  \urlprefix\url{http://link.aps.org/doi/10.1103/PhysRevLett.104.237401}.

\bibitem{gerlachPlasmaron74}
\bibinfo{author}{Gerlach, E.} \& \bibinfo{author}{Rautenberg, M.}
\newblock \bibinfo{title}{The {Dynamical} {Conductivity} for {Ionized}
  {Impurity} {Scattering}}.
\newblock \emph{\bibinfo{journal}{physica status solidi (b)}}
  \textbf{\bibinfo{volume}{65}}, \bibinfo{pages}{K13--K17}
  (\bibinfo{year}{1974}).
\newblock
  \urlprefix\url{http://onlinelibrary.wiley.com/doi/10.1002/pssb.2220650145/abstract}.

\bibitem{aliCava14}
\bibinfo{author}{Ali, M.~N.} \emph{et~al.}
\newblock \bibinfo{title}{The {Crystal} and {Electronic} {Structures} of
  {Cd}3as2, the {Three}-{Dimensional} {Electronic} {Analogue} of {Graphene}}.
\newblock \emph{\bibinfo{journal}{Inorganic Chemistry}}
  \textbf{\bibinfo{volume}{53}}, \bibinfo{pages}{4062--4067}
  (\bibinfo{year}{2014}).
\newblock \urlprefix\url{http://dx.doi.org/10.1021/ic403163d}.

\end{thebibliography}

\end{document}